\documentclass[twocolumn,showpacs,nofootinbib,longbibliography,notitlepage]{revtex4-2}
\usepackage{etex}
\usepackage{amsmath,amssymb,amsthm}
\usepackage{Physics}
\usepackage[colorlinks=true,citecolor=blue,urlcolor=blue]{hyperref}
\usepackage[pdftex]{graphicx}
\usepackage{times,txfonts}
\usepackage{braket}
\usepackage{color}
\usepackage{natbib}
\usepackage{amsmath,blkarray}
\usepackage{mathtools}
\usepackage{latexsym}
\usepackage{tabularx, booktabs}
\usepackage{graphics,epstopdf}
\usepackage{graphicx}
\usepackage{float}
\usepackage{graphicx}
\usepackage{amsfonts}
\usepackage{subcaption}
\usepackage{color,soul}

\newcommand{\be}{\begin{equation}}
	\newcommand{\ee}{\end{equation}}

\begin{document}
	\title{Device-independent certification of degeneracy-breaking measurements}
		\author{Prabuddha Roy}
	\email{prabuddharoy.94@gmail.com}
 \affiliation{National Institute of Technology Patna, Ashok Rajpath, Patna, Bihar 800005, India}
	\author{ Shyam Sundar Mahato}
 \affiliation{National Institute of Technology Patna, Ashok Rajpath, Patna, Bihar 800005, India}
	\author{ Sumit Mukherjee}
	\email{mukherjeesumit93@gmail.com}
 \affiliation{National Institute of Technology Patna, Ashok Rajpath, Patna, Bihar 800005, India}
	\author{A. K. Pan}
	\email{akp@phy.iith.ac.in}
	\affiliation{National Institute of Technology Patna, Ashok Rajpath, Patna, Bihar 800005, India}
 \affiliation{Department of Physics, Indian Institute of Technology Hyderabad, Telengana-502284, India }

	\begin{abstract}

In a device-independent Bell test, the devices are considered to be black boxes and the dimension of the system remains unspecified. The dichotomic observables involved in such a Bell test can be degenerate and one may invoke a suitable measurement scheme to lift the degeneracy. However, the standard Bell test cannot account for whether or up to what extent the degeneracy is lifted, as the effect of lifting the degeneracy can only be reflected in the post-measurement states, which the standard  Bell tests do not certify.  In this work, we demonstrate the device-independent certification of degeneracy-breaking measurement based on the sequential Bell test by multiple observers who perform degeneracy-breaking unsharp measurements characterized by positive-operator-valued measures (POVMs) - the noisy variants of projectors. The optimal quantum violation of Clauser-Horne-Shimony-Holt inequality by multiple sequential observers eventually enables us to certify up to what extent the degeneracy has been lifted. In particular, our protocol certifies the upper bound on the number of POVMs used for performing such measurements along with the entangled state and measurement observables. We use an elegant sum-of-squares approach that powers such certification of degeneracy-breaking measurements. 

	\end{abstract}
	\maketitle
	\section{Introduction}
	\label{I}

 In the standard \cite{home} quantum theory, the measurement process is interpreted as an entangling interaction between the states of the measuring apparatus and the observed system such that the latter is reduced to one of the eigenstates of the corresponding measurement operator associated with the measured observable - commonly known as the collapse of the wave function. While for non-degenerate observables the system indeed reduces to one of the eigenstate \cite{vonneumann}, for a degenerate observable, the eigenstate to which the system collapses depends on the prepared state on which the measurement is being performed. \cite{{luder1},{luder2}}. Note however that, for degenerate observables, there is provision to develop a suitable measurement scheme to lift the degeneracy so that the system can be made to collapse to a unique set of eigenstates depending upon the way the degeneracy is being lifted. To understand the degeneracy-breaking measurement scheme, in the following paragraph we briefly discuss the technical part of it.

{Let an operator $\hat{A}$ has $r$ number of discrete eigenvalues $\{a_r\}$ having degree of degeneracy $\{l_r\}$. Let $\{P_{r}^{i}=|\chi_{r}^{i}\rangle\langle\chi_{r}^{i}|\}$ with $i= 1,2,...,l_r $ is the set of rank-$1$ eigen-projectors of $\hat{A}$ satisfying $\sum_{i,r} P_{r}^{i}=\mathbb{I}_d$. Here $\mathbb{I}_d$ is the identity matrix in dimension of the Hilbert space $d=\sum_{r} [ \ l_{r}]$ where $[.]$ denotes the cardinality. Given a prepared state $\rho$, if the measurement process is degeneracy-preserving \cite{{luder1},{luder2}} then the reduced density matrix can be written as $\rho_{r}=\sum_{r} P_{r} \rho P_{r}$ where $	P_{r}=\sum_{i=1}^{l_{r}} P_{r}^{i}=\sum_{i=1}^{l_{r}}|\chi_{r}^{i}\rangle\langle\chi_{r}^{i}|$ and $|\chi_{r}^{i}\rangle$ is the eigenstate corresponding to the eigenvalue $a_{r}$ having degree of degeneracy $l_{r}$. In contrast, if measurement procedure is fully degeneracy-breaking,  the  reduced density matrix is $\rho^{\prime}_{r}=\sum_{r,i} P_{r}^{i} \rho P_{r}^{i}$. The degeneracy-breaking state reduction rule may be read as fine-graining of the degeneracy-preserving projection rule. There may be intermediate scenarios between fully degeneracy-breaking and degeneracy-preserving measurements which we call partially degeneracy-breaking measurements.}
	
It naturally follows from the preceding discussion that $\rho_{r}$ and $\rho^{\prime}_{r}$ are inequivalent unless the observable is non-degenerate.  For the measurement of degenerate observable, $\rho_{r}$ is expected to be less mixed than $\rho_{r}^{\prime}$, i.e., a larger amount of residual coherence remains in  $\rho_{r}$. Crucially, given a density matrix $\rho$, the expectation value of any arbitrary observable $\langle \hat{A}\rangle=\Tr[\hat{A}\rho]$ remains the \textit{same}, irrespective of the degeneracy-breaking or -preserving measurement is implemented. The measurement statistics of $\hat{A}$ only pertain to the probabilities of the different outcomes (eigenvalues) and is \textit{not} concerned about the post-measurement states. However, the scenario becomes interesting if a sequential measurement of another observable is performed posterior to $\hat{A}$. In such a case, different post-measurement states may provide different statistics in sequential measurements. 

The aim of this work is to certify the degeneracy-breaking measurement through the quantum violation of Bell's inequality. {Bell's inequalities provide an avenue to test the eccentric non-classical feature widely known as quantum nonlocality in the device-independent (DI) way. In a DI Bell test,  the dimension of the system remains unspecified and the devices are considered to be back-boxes, i.e., the inner working of the devices remains uncharacterized. Only the observed output statistics are
enough to certify non-locality.} Such DI certification of quantum correlations has recently found a plethora of applications in quantum information theory \cite{brunnerrev,bar05,acin06,acin07,pir09}. For the certification of degeneracy-breaking measurement, we consider the Clauser-Horn-Shimony-Holt (CHSH) inequality \cite{chsh} - the simplest form of Bell's inequality involving two parties, two measurements per party and two outcomes per measurement (2-2-2) scenario \cite{brunnerrev}. Such certification has not hitherto been explored and hence we first explain the motivation of the present work a bit elaborately in the following section.

This paper is organized as follows. First, before we present our main result, we briefly discuss the motivation (Sec. \ref{moti}) of our work.  In Sec. \ref{sec2}, we recapitulate the essence of the CHSH inequality by explicitly providing the DI self-testing based on the optimal quantum violation by introducing an elegant sum-of-squares (SOS) approach recently developed in \cite{sneha}. Next, in Sec. \ref{sec3}, we revisit the sequential independent sharing \cite{asmita2019,silva2015,sasmal2018,zhang21,brown2020} of CHSH-nonlocality in the DI way. In Sec. \ref{sec4}, we first provide a detailed analysis of the degeneracy-breaking measurement scheme. Subsequently, by introducing different $g$-POVMs, we provide the analytical treatment of quantum bound of the CHSH expressions between Alice-Bob$_{2}$ in terms of the unsharpness parameter in Secs. \ref{sec4a} and \ref{sec4b} and discuss the certification argument in Sec. \ref{sec4c}. Sec. \ref{sec5} is equipped with the way the certification of $g$-POVMs has been done using the certified unsharpness parameter of Bob$_{1}$. Finally, we summarize our results in Sec. \ref{sec6} and provide future directions.
	
\section{Motivation of the present work}\label{moti}
	
 It is known that the optimal quantum violation of the CHSH inequality self-tests \cite{mayers} the local observables of both Alice and Bob as well as the state shared between these two parties, \textit{viz.,} the observables for both the parties are anti-commuting and the state shared between them to be maximally entangled. Since in Bell test the devices of both the parties remain uncharacterized and the dimension of the system remains unspecified, the nonlocal correlation is self-tested in DI way  \cite{mayers,supicrev,sarkar,tavakoli19a,smania}.

It is essential to note here that the quantum optimal violation of the CHSH inequality remains the \textit{same} even if Alice and Bob share $m$ copies of the two-qubit maximally entangled state, as long as the local observables in dimension $d=2^{ m}$ are dichotomic. If the local dimension $d>2$, the observable is degenerate and there are many possible ways of implementing the quantum measurement depending on how the degeneracy is being lifted. To put it succinctly, let us introduce the notion of $g$-projector measurements and for $g=2$ the measurement is degeneracy-preserving. For $g>2$, the measurement is degeneracy-breaking which can be seen as a kind of fine-graining measurement and $g=d$ corresponds to the fully degeneracy-breaking measurement. For such a fully degeneracy-breaking measurement, the system can reduce to one of the eigenstates corresponding to a unique set of eigenfunctions fixed by the degeneracy-breaking scheme. For instance, if $d=4$ and $g=4$ then the $4$-projector fine-graining measurement is fully degeneracy-breaking. Similarly, in this case, $3$- and $2$-projector measurements are partially degeneracy-breaking and degeneracy-preserving measurements, respectively.
	 
\begin{figure}[ht]
\centering 
{{\includegraphics[width=0.9\linewidth]{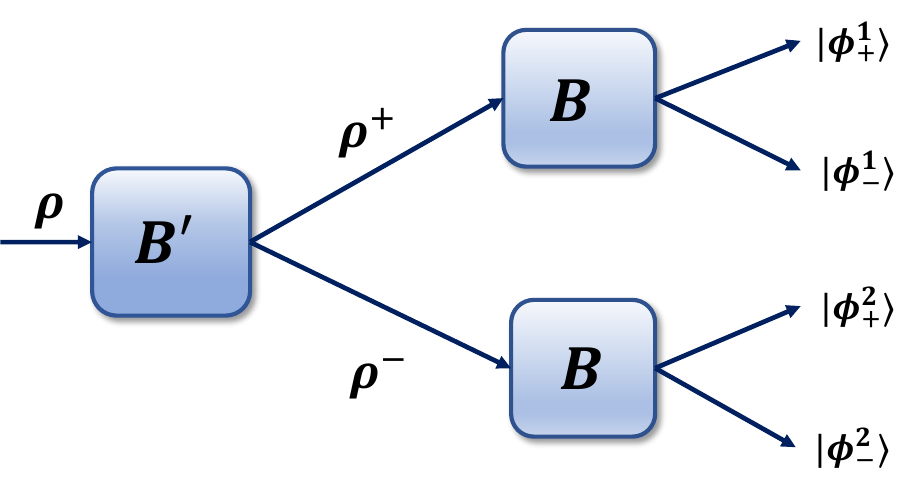}}}
\caption{ A schematic diagram shows an example of degeneracy-breaking measurement for $g=4$. Bob$_{1}$ poses a  two-qubit dichotomic observable  $B$. In order to break the degeneracy, Bob$_{1}$ introduces another dichotomic observable $B^{\prime}$ prior to $B$ which follows $\qty[B,B^{\prime}]=0$. Given a state $\rho$, Bob$_{1}$ first performs the measurement $B^{\prime}$ and partially breaks the degeneracy to one of the non-unique eigensubspaces. Subsequently, Bob$_1$ measures $B$ on one of those reduced eigensubspaces. Finally, the system collapses to one of the four eigenstates.  }\label{db}
\end{figure}
{To provide more clarity let us introduce a brief sketch of how degeneracy-breaking measurements can be performed for a local two-qubit ($d=4$) system as depicted in FIG. \ref{db}. Let us assume an observable   $B=P_{+} - P_{-}=\ket{\phi^{1}_{+}}\bra{\phi^{1}_{+}}+\ket{\phi^{2}_{+}}\bra{\phi^{2}_{+}}-\ket{\phi^{1}_{-}}\bra{\phi^{1}_{-}}-\ket{\phi^{2}_{-}}\bra{\phi^{2}_{-}}$ where $P_{+}=\ket{\phi^{1}_{+}}\bra{\phi^{1}_{+}}+\ket{\phi^{2}_{+}}\bra{\phi^{2}_{+}}$ and $p_{-}=\ket{\phi^{1}_{-}}\bra{\phi^{1}_{-}}-\ket{\phi^{2}_{-}}\bra{\phi^{2}_{-}}$ are the rank-2 eigenprojectors having eigenvalues $+1$ and $-1$ respectively. Here $\ket{\phi^{1}_{+}}\bra{\phi^{1}_{+}}$ and $\ket{\phi^{2}_{+}}\bra{\phi^{2}_{+}}$ are the two eigenstates of $\hat{B}$ having the same eigenvalue $+1$ and similarly for the other two eigenstates.  In order to break the degeneracy one needs to introduce another observable $B^{\prime}$ which is commuting to $B$ and hence posses common eigenstates. Let us assume that $B^{\prime}=\ket{\phi^{1}_{+}}\bra{\phi^{1}_{+}}+\ket{\phi^{1}_{-}}\bra{\phi^{1}_{-}}-\ket{\phi^{2}_{+}}\bra{\phi^{2}_{+}}-\ket{\phi^{2}_{-}}\bra{\phi^{2}_{-}}$. Here $\ket{\phi^{1}_{+}}$ and $\ket{\phi^{1}_{-}}$ are the eigenstates of $B^{\prime}$ corresponding to the eigenvalue $+1$ of $B^{\prime}$, and $ \ket{\phi^{2}_{+}}$ and $\ket{\phi^{2}_{-}}$ are the  eigenstates with eigencalue $-1$.  To lift the degeneracy, one first performs the non-selective measurement of  $B^{\prime}$ on the given state $\rho$ and then applies $B$ sequentially. The system then collapses to one of the four eigenstates. 

For our purpose, we shall be shortly introducing the POVMs which are noisy variant of the projectors. Now, for $d=4$ the POVMs corresponding to the projectors are $\{E_{b}^{i}\}$ with $i \in \{1,2\}$ and $b\in \{-1,+1\}$. In case of fully degeneracy-breaking measurement ($g=4$), to obtain the post-measurement states the Kraus operators are constructed corresponding to the POVMs $\{E_{+}^{1},E_{+}^{2},E_{-}^{1},E_{-}^{2}\}$ whereas in case of partially degeneracy-breaking measurement ($g=3$), the Kraus operators are constructed corresponding to the POVMs $\{E_{+}^{12}\equiv E_{+}=(E_{+}^{1}+E_{+}^{2}), E_{-}^{1}$, $E_{-}^{2}\}$ or  \{$E_{+}^{1}$, $E_{+}^{2}, E_{-}^{12}\equiv E_{-}=(E_{-}^{1}+E_{-}^{2})\}$. For degeneracy-preserving measurement ($g=2$), the Kraus operators are constructed corresponding to the $g$-POVMs $\{E_{+}=(E_{+}^{1}+E_{+}^{2}), E_{-}=(E_{-}^{1}+E_{-}^{2})\}$. }
	  
We note again that in a Bell experiment the degeneracy-breaking and -preserving measurements cannot be distinguished from the measurement statistics and the quantum optimal value is independent of the dimension $d$. It then immediately follows that the standard Bell experiment cannot distinguish which measurement scheme (the value of $g$) is implemented to obtain the statistics. Therefore, the optimal quantum violation of a conventional Bell test neither certifies the post-measurement states nor the number of projectors ($g$) used to realize the measurements. In this work, we show that the sequential Bell test possesses the potential to certify the post-measurement states. We provide a  DI certification protocol that certifies the $g$-projector measurement through the sequential quantum violation of the CHSH inequality. Although a number of works have been reported \cite{sala,Mayato,Kathakali,Sudheer,budroni14,luo,asmita21} aiming to distinguish (not to certify) degeneracy-preserving and fully degeneracy-breaking measurement, it is fair to say that they are quite far from the motivation of the present work.
	
We consider the sequential CHSH scenario comprises of single Alice and an arbitrary $k$ number of independent sequential Bobs (Bob$_{k}$). While our protocol works for dichotomic observables in arbitrary dimension $d>2$, for convenience we consider that Alice and Bob$_{1}$ share $m$-copies of two-qubit entangled states. If Bob$_{1}$ performs projective measurement, degeneracy-breaking or not, the entanglement will be completely destroyed; thereby, quantum violation of the CHSH inequality between Alice and Bob$_{2}$ cannot be achieved. To ensure the CHSH violation between Alice and Bob$_{2}$, Bob$_{1}$ must have to perform the unsharp measurements, i.e., POVMs \cite{busch}. Here we consider those POVMs which are the noisy variants of sharp projective measurements. If Bob$_{1}$ performs POVMs, a residual entanglement may remain which can be used by Alice and Bob$_{2}$ to demonstrate the violation of CHSH inequality, and the chain continues for Bob$_{k}$. It is important to note here that there will be a trade-off between the quantum violations of the CHSH inequality between Alice and sequential Bobs depending upon the unsharpness of the measurement instrument used by Bob$_k$ and the value of $g$. For example, one may take the minimum value of the unsharpness parameter for Bob$_1$ so that the CHSH inequality is just violated. In such a case, the entanglement state will be minimally disturbed and subsequently, there will be a high possibility of obtaining a CHSH violation between Alice and Bob$_2$.  The process continues until $k^{th}$ sequential Bob gets the quantum violation of CHSH inequality.
	
Importantly, for $g$-POVM measurements, the disturbance created by Bob$_1$ is also dependent on the value of $g$. For $g=2$, the residual entanglement after Bob$_{1}$'s measurement is higher than $g>2$. Hence, there is a scope to demonstrate the sharing of quantum correlations to more observers if $2$-POVM measurements are performed. By observing the quantum violations of CHSH inequality by Alice and sequential Bobs, we demonstrate DI certification of the number $(g)$ of POVM used to implement the measurement. Therefore, our protocol self-tests the state, observables, and the degeneracy-breaking or -preserving measurement solely from the statistics.

	\section{Optimal quantum violation of the CHSH inequality}\label{sec2}
	
{The standard CHSH experiment involves two space-like separated parties (say, Alice and Bob) who share systems having a common past. Alice (Bob) performs local measurements on her (his) subsystem upon receiving inputs  $x\in \{1,2\}$  $(y \in \{1,2\})$, and returns outputs $a \in \{-1,+1\}$ $(b \in \{-1,+1\})$. Representing $A_{x}$ and $B_{y}$ as respective observables of Alice and Bob, the CHSH expression is given by
\begin{equation}
		\label{bellor}
		\mathcal{B}=\qty(A_{1}+A_{2}) B_{1} +\qty(A_{1}-A_{2}) B_{2}.
	\end{equation}

where $\langle A_{x}B_{y}\rangle=\underset{a,b=\pm1}{\sum} ab\ p(a,b|A_{x}B_{y})$} and $p(a,b|A_{x}B_{y})$ is the joint probability distribution. For any local realist theory, the expectation value $\langle \mathcal{B} \rangle$ of the CHSH expression $\mathcal{B}$ is bounded by $\langle \mathcal{B} \rangle\leq2$. 
In quantum theory, $\langle \mathcal{B} \rangle_{Q}=\Tr[\mathcal{B}\ \rho_{AB}]$ where $\rho_{AB}$ is the joint quantum state of Alice and Bob.  The optimal quantum value of the CHSH expression has shown to be $\langle \mathcal{B} \rangle^{opt}_{Q}=2\sqrt{2}$ \cite{cirelson}. The optimal value is achieved when the observables local to Alice and Bob are anti-commuting and the shared state $\rho_{AB}$ is a maximally entangled state.
	
By using an elegant SOS approach we explicitly show in the Appendix \ref{appsos} that the optimal value $\langle \mathcal{B} \rangle^{opt}_Q$ remains the same even if Alice and Bob perform dichotomic measurements on a shared state $\rho_{AB} \in \mathcal{H}_{A}^{d}\otimes \ \mathcal{H}_{B}^{d}$ where $d\geq 2$ is arbitrary. Thus if the measurement is done on such a higher dimensional entangled state with $d=2^{m}$ ($m$ being an integer), an observer (say, Bob) gets the provision to implement his measurement in different ways \textit{viz.,} degeneracy-preserving, partially degeneracy-breaking, or fully degeneracy-braking. One may then ask whether the different degrees of degeneracy-breaking measurements can be certified through the quantum violation of CHSH inequality. 

For example, if Bob's local subsystem is two-qubit $(d=4)$, Bob can choose to measure the dichotomic observables as $B_{y}= P_{+|y}-P_{-|y}$ with $ y \in \{1,2\}$ in the degeneracy-preserving ($g=2$) scheme where $P_{\pm|y}=\frac{1}{2}\Big(\mathbb{I}\pm B_{y}\Big)$ are the rank-2 projectors or can implement degeneracy-breaking $(g=4)$ scheme by breaking each rank-2 projector into rank-1 projectors i.e., $P_{+|y}=P^{1}_{+|y}+P^{2}_{+|y}$ ; $P^{-}_{y}=P^{1}_{-|y}+P^{2}_{-|y}$ (explicitly shown for $\langle\mathcal{B}^{2}\rangle^{g=4}_Q$ in Appendix \ref{appg4}). One may also consider three projector $(g=3)$ measurements. It is straightforward to show that irrespective of the measurement scheme implemented, the value of each correlation $\langle A_{x}B_{y}\rangle$ and further the CHSH expression $\qty(\Tr[\rho_{AB} \mathcal{B}])$ between Alice and Bob$_{1}$ remains the same. This argument holds good for any arbitrary dimension and arbitrary $g$ value. However, the post-measurement states for different $g$ values can be different. This can be captured through the sequential quantum violation of CHSH inequality which in turn enables us to certify the degree of degeneracy-breaking measurement $g$.

\section{DI Sharing of nonlocality in the CHSH scenario for degeneracy-preserving measurement}\label{sec3}
We start by noting that the sequential sharing of nonlocality was first introduced in \cite{silva2015} to demonstrate that two independent sequential observers on one side can share the nonlocality given the shared entangled system is of minimum dimension, i.e., maximally entangled two-qubit state. In a recent work, \cite{brown2020}, by using a two-qubit entangled state it has been shown that arbitrary independent Bobs can sequentially share nonlocality by adjusting the trade-off between measurement settings and unsharpness parameters.
		
Here, we also consider the sequential sharing of nonlocality but without assuming the dimension of the system - a key element for DI certification. While the purpose of the work in \cite{brown2020} was to demonstrate the sharing of nonlocality by an arbitrary number of independent observers using a minimum dimensional system, we demonstrate the sharing of nonlocality in a DI way without assuming the dimension of the system. Such DI way of sharing nonlocality has not hitherto been demonstrated. Note again that the aim of the present work is to demonstrate the DI certification of degeneracy-breaking measurement. For achieving our purpose, we invoke the sequential quantum violation of Bell's inequality as a tool. This is because the certification of degeneracy-breaking measurement necessitates the DI certification of the post-measurement state. Thus, consideration of two independent sequential Bobs suffices our requirement of ours. Hence, in this section, we first demonstrate the sequential sharing of nonlocality in the aforementioned scenario without assuming the dimension of the shared system as well as the inner working of the measurement devices of Alice and Bobs. In this section, we will discuss certification of degeneracy-preserving measurement i.e., $g=2$ case.
	
The sequential Bell experiment comprises one Alice who performs sharp measurement, and an arbitrary $k$ number of independent Bobs (Bob$_{k}$) who perform unsharp measurements with respective unsharpness parameters $\lambda_{k}$ sequentially. After Bob$_{k}$'s measurement, he relays his subsystem to Bob$_{(k+1)}$. Note that since all Bobs are independent of each other, each Bob does not have any information about previous Bobs measurements as well as outcomes. Now, here we use the quantum violation of CHSH inequality \cite{chsh} for the purpose of sequential sharing of nonlocality. In order to obtain the reduced state, averaged over Bob$_{k}$'s measurements and outcomes, between Alice and Bob$_{(k+1)}$, let us first characterize Bob's measurement by Kraus operators $\{K_{b_{k}|y_{k}}\}$, where $\forall k$, $b_{k}=\pm 1$,  $\sum_{b_{k}} K_{b_{k}|y_{k}}^{\dagger}K_{b_{k}|y_{k}}=\mathbb{I}$ and $K_{b_{k}|y_{k}}= \sqrt{E_{b_{k}|y_{k}} }$. Here $E_{b_{k}|y_{k}} = \frac{1}{2}\left(\mathbb{I}+\lambda_{k}{b_{k}} B_{y_{k}}\right)$ are the POVMs corresponding to observables $ B_{y_{k}}$ satisfying $\sum_{b_{k}}E_{b_{k}|y_{k}}=\mathbb{I}$ \cite{busch,kunj}. The Kraus operators can be written as	
\begin{eqnarray}\label{kraus}
K{\pm|y_{k}}&=& \sqrt{\frac{(1\pm\lambda_{k})}{2}}P_{+|y_{k}} +\sqrt{\frac{(1\mp\lambda_{k})}{2}}P_{-|y_{k}}\nonumber\\
& \equiv& \alpha_{k} \mathbb{I} \pm \beta_{k} B_{y_{k}} 
\end{eqnarray} 
where $P_{\pm|y_{k}}$ are the projectors corresponding to each of the eigenvectors of $B_{y_{k}}$ and 
\begin{eqnarray} \label{albe}
\alpha_{k} &=& \dfrac{1}{2}\left[\sqrt{\dfrac{(1+\lambda_{k})}{2}}+ \sqrt{\dfrac{(1-\lambda_{k})}{2}}\ \right];\nonumber\\
\beta_{k} &=& \dfrac{1}{2}\left[\sqrt{\dfrac{(1+\lambda_{k})}{2}}-\sqrt{\dfrac{(1-\lambda_{k})}{2}}\ \right],
\end{eqnarray}
satisfying $\alpha_{k}^2 +\beta_{k}^2 = 1/2$. We take the same unsharpness parameter $\lambda_{k}$ for the two measurements of Bob$_{k}$, as the unsharpness is the property of the device, irrespective of which observable is being measured.	Then, after $k^{\text{th}}$ Bob's measurement, the average reduced state shared between Alice and Bob$_{k+1}$ is given by

\begin{figure*}
\centering
\includegraphics[scale=0.4]{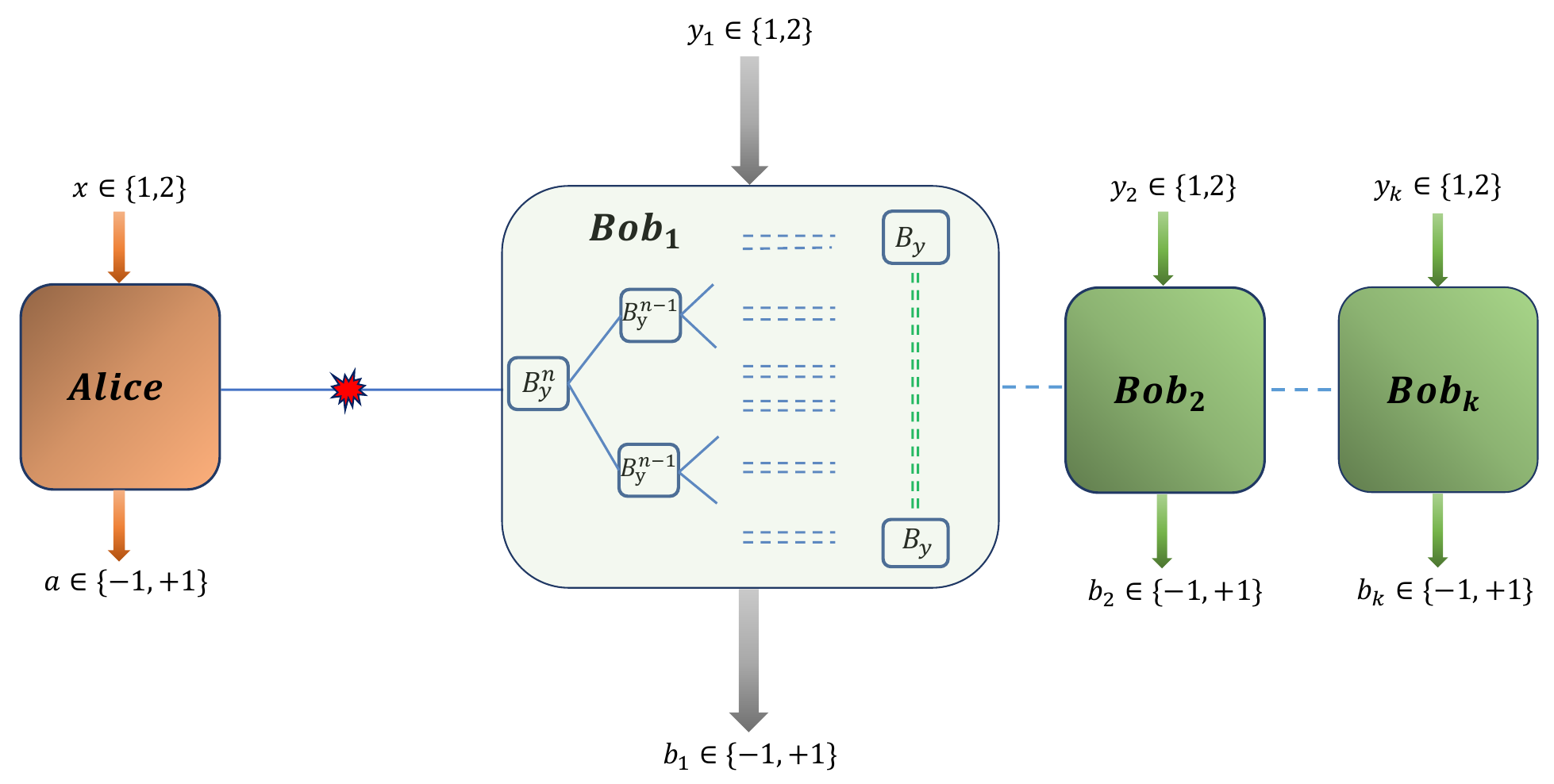}
\caption{ Schematic diagram for our proposed variant of sequential Bell test consisting of one Alice and multiple independent Bobs (Bob$_{k}$). By `independent' we mean that Bob$_2$ does not have any information about Bob$_1$'s measurement and so on. Bob$_{1}$ implements degeneracy-preserving(breaking) measurement on his local subsystem by introducing noisy POVMs and sends the post-measurement state to the next sequential independent Bob. }
\end{figure*}

\begin{eqnarray}
\rho_{AB_{(k+1)}}&=&\dfrac{1}{2} \sum_{b_{k}\in \{ +,-\}}^{}\sum_{y_{k}=1}^{2} \left(\mathbb{I} \otimes K_{b_{k}|y_{k}} \right) \  \rho_{AB_{k}}  \left(\mathbb{I} \otimes K_{b_{k}|y_{k}} \right) \label{g2}.
\end{eqnarray}

{Now, we are in a position to demonstrate DI sequential sharing of nonlocality between Alice-Bob$_{k}$. We denote $\langle\mathcal{B}^{k}_{Q}\rangle^{g=2}$ by the quantum value of the Bell expression between Alice and Bob$_{k}$. By using Eq. (\ref{kraus}), the quantum value of the CHSH expression $\langle \mathcal{B}^{1} \rangle_Q^{g=2}$  corresponding to Alice and Bob$_{1}$ is given by 
\begin{equation}\label{bg1}
\langle \mathcal{B}^{1} \rangle_Q^{g=2}=\Tr[\rho_{AB_{1}} \mathcal{B}]=\lambda_{1}\langle 	\mathcal{B}\rangle_Q^{opt}	\end{equation}
where $\rho_{AB_{1}}\in \mathcal{H}_{A}^{d}\otimes \mathcal{H}_{B_{1}}^{d}$ is the bipartite state shared between Alice and Bob$_{1}$ and $\lambda_{1}$ is the unsharpness parameter of Bob$_{1}$'s instrument. As mentioned earlier that $\langle \mathcal{B}^{1}\rangle_Q$ remains the same irrespective of the way measurement is implemented by Bob$_{1}$, i.e., $\langle \mathcal{B}^{1}\rangle_Q^{g}\equiv \langle \mathcal{B}^{1}\rangle_Q$.}
	 
After the measurement of Bob$_{1}$, the average state shared between Alice and Bob$_{2}$ is given by
\begin{eqnarray}\label{lu2}
\rho_{AB_{2}}&=&\dfrac{1}{2} \sum_{b_1 \in \{ +,-\}}\sum_{y_{1}=1}^{2} \left(\mathbb{I} \otimes K_{b_{1}|y_{1}}\right)  \rho_{AB_{1}}  \left(\mathbb{I} \otimes K_{b_{1}|y_{1}}\right)\nonumber\\
&=&2\alpha_2^2 \  \rho_{AB_{1}} + \beta_2^2 \sum_{y_{1}=1}^{2} \qty(\mathbb{I} \otimes B_{y_{1}}) \ \rho_{AB_{1}}\qty(\mathbb{I} \otimes B_{y_{1}}).
\end{eqnarray}

By using Eq. (\ref{lu2}), the quantum value of the CHSH expression for Alice and Bob$_{2}$ is evaluated as (see appendix \ref{appsos2} for detailed derivation) 
 \begin{eqnarray}\label{b2bell}
\langle \mathcal{B}^{2} \rangle_Q^{g=2}&=& \underset{\rho_{AB_{1}},\{A_{x}\},\{B_{y}\}}{\max}\qty(\Tr[\rho_{AB_{2}} \mathcal{B}])\\
&=&\underset{\rho_{AB_{1}},\{A_{x}\},\{B_{y}\}}{\max}\qty(\Tr[\rho_{AB_{1}}\qty((A_{1}+A_{2})B_{1}^{\prime}+(A_{1}-A_{2})B_{2}^{\prime})]) \nonumber
\end{eqnarray}

Naturally, the maximum value can be obtained when Bob$_{2}$ performs sharp measurement.
The explicit forms of $B_{1}^{\prime}$ and $B_{2}^{\prime}$ are derived as	
	
  \begin{eqnarray}\label{primeobs}
  \nonumber
B_{1}^{\prime}&=&(2\alpha_{2}^2+\beta_{2}^2)B_{1}+\beta_{2}^{2}B_{2}B_{1}B_{2}\\
B_{2}^{\prime}&=&(2\alpha_{2}^2+\beta_{2}^2)B_{2}+\beta_{2}^{2}B_{1}B_{2}B_{1}.
  \end{eqnarray}

In Appendix \ref{appobser}, we prove that the quantum values  $\langle \mathcal{B}^{1} \rangle_Q^{g=2}$ and $\langle \mathcal{B}^{2} \rangle_Q^{g=2}$ simultaneously optimised when both Bob$_{1}$'s as well as Bob$_{2}$'s observables are anti-commuting. We also prove that to obtain the maximum quantum value of $\langle\mathcal{B}^{2}\rangle_{Q}^{g=2}$ the choices of observables of Bob$_{2}$ have to be the same as Bob$_{1}$. Note that, Eq. (\ref{b2bell}) has a similar form of CHSH expression as in Eq. (\ref{bellor}). However, here $(B_{1}^{\prime})^{2}\neq \mathbb{I}$ and $(B_{2}^{\prime})^{2}\neq \mathbb{I}$, and hence they need to be properly normalized.  Hence, in order to obtain the maximum quantum value of Eq. (\ref{b2bell}), considering $\omega_{1} =||\left(A_{1}+A_{2}\right)|\psi\rangle_{AB}||_{2}$, $\omega_{1}^{\prime} =|| B_{1}^{\prime}|\psi\rangle_{AB} ||_{2}$, $\omega_{2} =||\left(A_{1}-A_{2}\right)|\psi\rangle_{AB}||_{2}$ and $\omega_{2}^{\prime} =|| B_{2}^{\prime}|\psi\rangle_{AB}||_{2}$, and by using the SOS approach \cite{pan2020} we derive ( see Appendix \ref{appsos2} for detailed derivation)
\begin{eqnarray}\label{maxchsh}
\langle \mathcal{B}^{2}\rangle_Q^{g=2}= \max \qty(\omega_{1}\omega_{1}^{\prime} +\omega_{2}\omega_{2}^{\prime}).
\end{eqnarray}

As we have already proved in Appendix \ref{appsos}, to obtain the optimal quantum value Alice's and Bob's observables have to be mutually anticommuting. Hence, for Bob's (unnormalized) observables $B_{1}^{\prime}$ and $B_{2}^{\prime}$ we require,
\begin{eqnarray}
\label{anti}
\{B_{1}^{\prime},B_{2}^{\prime}\}=4\alpha_{2}^2(\alpha_{2}^2+2\beta_{2}^{2}) \{B_{1},B_{2}\}+\beta_{2}^{4}\{B_{1},B_{2}\}^{3}=0 .
\end{eqnarray}
Since $\alpha_{2}>0$ and $\beta_{2}\geq 0$, Eq. (\ref{anti}) gives $\{B_{1},B_{2}\}=0$. In other words, the observables of Bob$_{2}$ have to be anticommuting to obtain the maximum quantum value of the Bell expression $\langle \mathcal{B}^{2} \rangle_Q^{g=2}$. This, in turn, provides 
\begin{eqnarray}\label{otilde}
\omega_{1}^{\prime}=\sqrt{\qty(2\alpha_{2}^2)^{2}+\qty(2\alpha_{2}^2+\beta_{1}^2) \ \beta_{2}^{2} \{B_{1},B_{2}\}^{2}}=2\alpha_{2}^2,
\end{eqnarray}
and $\omega_{2}^{\prime}=\omega_{1}^{\prime}=2\alpha_{2}^2$. By putting $\omega_{1}^{\prime},\omega_{2}^{\prime}$ from Eq. (\ref{otilde}) we have
\begin{eqnarray}\nonumber
\langle \mathcal{B}^{2} \rangle_Q^{g=2}= 2\alpha_{2}^2 \ \max\Big(\omega_{1}+\omega_{2}\Big)=2\alpha_{2}^2\langle \mathcal{B} \rangle_{Q}^{opt}.
\end{eqnarray}
where we used Eq. (\ref{optbnn}).
Putting the value of $\alpha_{2}$, we have 	$\langle\mathcal{B}^{2}\rangle^{g=2}_Q$ in terms of unsharpness parameter as
\begin{equation}\label{llbell1}
\langle \mathcal{B}^{2} \rangle_Q^{g=2}=\dfrac{1}{2}\Big(1+\sqrt{1-\lambda_{1}^2}\Big)\langle \mathcal{B} \rangle_Q^{opt}
\end{equation}

{Since the optimal quantum value of $\langle \mathcal{B} \rangle_Q^{opt}$ is $2\sqrt{2}$, the lower bound of unsharpness parameter for Bob$_{1}$ is required for violating the CHSH inequality is calculated from Eq. (\ref{bg1}) as $\lambda_{1}=1/\sqrt{2}\approx 0.707$.
Now, in order to have a simultaneous quantum violation of CHSH inequality for Bob$_{1}$ as well as Bob$_{2}$, one requires $\langle \mathcal{B}^{1}\rangle_Q>2$ and $\langle \mathcal{B}^{2} \rangle_Q^{g=2}>2$. This demands Bob$_{1}$'s measurements to be necessarily unsharp one. The upper bound to $\lambda_{1}$ is calculated by putting  $\lambda_{2}=1$, in Eq. (\ref{llbell1}) when Bob$_{2}$ just surpasses the classical bound and is given by, $(\lambda_{1})_{\max}= \sqrt{2\qty(\sqrt{2}-1)}\approx 0.912$. If the values of $\langle \mathcal{B}^{1} \rangle_Q$ and $\langle \mathcal{B}^{2} \rangle_Q^{g=2}$ are experimentally realized one can  fix the range of unsharpness parameter of Bob$_{1}$'s measurement instrument to be $\lambda_{1}\in [0.707,0.912]$. As the certification of degeneracy-breaking measurements demands the quantum violation of CHSH inequality by at least two observers, given any value of $\lambda_{1}$ within the above range enables the certification.}  

We note here that the semi-DI certification of post-measurement qubit states in the prepare-measure scenario was studied in \cite {mohan,paw20,sumit} and experimentally tested \cite{anwar2020,foletto} . We considered DI self-testing of the post-measurement state in any arbitrary dimension in the entangled scenario.   
	
\section{Certification of Degeneracy-breaking measurement}\label{sec4}
To certify the degree $(g)$ of degeneracy-breaking, we invoke the sequential quantum violation of CHSH inequality by multiple sequential observers. Now, in the sequential scenario, Bob$_{1}$ has the freedom to implement his measurement by performing $g$-POVMs - the noisy variants of $g$-projectors. As mentioned earlier in Sec. \ref{I}, $g=2$ corresponds to standard degeneracy-preserving measurement. For $g>2$, the measurement is degeneracy-breaking and for $g=d$ the measurement is fully degeneracy-breaking. It is crucial to note here that if Bob$_1$ implements his measurement by considering the number of POVMs $g>2$, the post-measurement states are different than that of $g=2$ case. Intuitively, for $g>2$ case, the initial state is expected to be more disturbed than $g=2$ case. It is then  obvious that the quantum value of the CHSH expression for Alice and Bob$_{2}$ will be smaller in $g>2$ than that in the case of $g=2$. The higher the value of $g$, lower the quantum violation of the CHSH inequality for Alice and Bob$_{2}$. 

To add more clarity, for example, let Alice and Bob$_{1}$ share a pair (i.e., $m=2$) of two-qubit entangled state $\rho_{AB_{1}}$ and hence the dimension of each of the local systems is four. Note that the standard sequential sharing scenario as discussed in Sec. \ref{sec3} corresponds to the standard non-selective degeneracy-preserving measurement scheme ( i.e., $g=2$).  In this case, recalling Eq. (\ref{g2}) the average state between Alice and Bob$_{2}$ is written as
\begin{equation}\label{rhoabg2}
\rho^{g=2}_{AB_{2}}=\frac{1}{2}\sum_{y_{1}=1}^{2}\sum_{b_{1}=\pm }\qty(\mathbb{I}\otimes P_{b_{1}|y_{1}}) \ \rho_{AB_{1}} \ \qty(\mathbb{I}\otimes P_{b_{1}|y_{1}})   
\end{equation}
where $P_{b_{1}|y_{1}}=\sum_{i=1}^{2}\ket{\phi^{i}_{b_{1}|y_{1}}}\bra{\phi^{i}_{b_{1}|y_{1}}}$ and $b_{1}=\pm 1$ are the eigenvalues of $B_{y_{1}}$. Bob$_{1}$'s dichotomic observable can be written as  $B_{y_{1}}=P_{+|y_{1}}- P_{-|y_{1}}$, where the rank-2 projectors are $P_{+|y_{1}}=\ket{\phi^{1}_{+|y_{1}}}\bra{\phi^{1}_{+|y_{1}}}+\ket{\phi^{2}_{+|y_{1}}}\bra{\phi^{2}_{+|y_{1}}}$ and  $P_{-|y_{1}}=\ket{\phi^{1}_{-|y_{1}}}\bra{\phi^{1}_{-|y_{1}}}+\ket{\phi^{2}_{-|y_{1}}}\bra{\phi^{2}_{-|y_{1}}}$. 
	
However, instead of implementing degeneracy-preserving measurement ($g=2$), Bob$_1$ can also invoke fully degeneracy-breaking measurement ($g=4$) as we consider Bob$_{1}$ system is two-qubit as an example.  {The reduced state after Bob$_{1}$'s measurement is given by  
\begin{equation}\label{rhoabg4}
\rho^{g=4}_{AB_{2}}=\frac{1}{2} \sum_{y_{1},i=1}^{2} \sum_{b_{1}=\pm }  \qty(\mathbb{I}\otimes P_{b_{1}|y_{1}}^{i}) \ \rho_{AB_{1}}\qty(\mathbb{I}\otimes P_{b_{1}|y_{1}}^{i})
\end{equation}}
where $P_{b_{1}|y_{1}}^{i}=\ket{\phi^{i}_{b_{1}|y_{1}}}\bra{\phi^{i}_{b_{1}|y_{1}}}$ (with $i=1,2$) are the rank-1 orthogonal projectors satisfying $\sum_{i}\sum_{b_{1}}P_{b_{1}|y_{1}}^{i}=\mathbb{I}$.

If Bob$_{1}$ performs the unsharp measurements then for degeneracy-preserving (or, fully degeneracy-breaking) measurements he uses suitable  POVMs which are the noisy variants of projectors. Then the maximum number of POVMs is limited to the local dimension of the system. Since the degeneracy-breaking measurement is more disturbing than the degeneracy-preserving one, the entanglement content of $\rho^{g>2}_{AB_{2}}$ is expected to be less than $\rho^{g=2}_{AB_{2}}$.  
	
We certify degeneracy-breaking or -preserving measurement for dichotomic observable of dimension $d>2$ through the sequential violation of the CHSH inequality. 	
Intuitively, it is expected that the higher the value of $g$ smaller the quantum value of the CHSH expression between Alice and Bob$_{2}$. We show that this feature plays a crucial role in certifying the lower bound of $g$ from the optimal quantum value of $\langle \mathcal{B}^{2} \rangle_Q^{g}$.

\subsection{Degeneracy-breaking measurement for $g=4$ case}\label{sec4a}
As already mentioned in the preceding Sec. \ref{sec4}, in the sequential scenario, Bob$_{1}$ performs non-selective unsharp measurement on his local system by considering any value of $g\geq 2$, and relaying the system to Bob$_{2}$ who performs sharp measurement. The maximum value of $g$ is the dimension $d$ of Bob$_{2}$'s local system, i.e.,  $g_{\max}=d$. Depending upon the value of $g$, measurements can be fully or partially degeneracy-breaking or degeneracy-preserving. For demonstration, let us assume the dimensions of the local systems are $d=2^{m}$ where $m$ is arbitrary and unknown. 
	
Let us explicitly consider the case of $g=4$. In this case, Bob$_{1}$ performs his measurements by considering four POVMs that can be constructed as  
\begin{eqnarray}\label{povm2q}
E_{+|y_{1}}^{1}&=& \dfrac{1}{4} \left[ \mathbb{I} + \lambda_{1} \left( B_{{y}_{1}}+ M_{y_{1}}\right)\right];\ 
E_{+|y_{1}}^{2}= \dfrac{1}{4} \left[ \mathbb{I} + \lambda_{1} \left( B_{{y}_{1}}- M_{y_{1}}\right)\right]\nonumber\\
E_{-|y_{1}}^{1}&=& \dfrac{1}{4} \left[ \mathbb{I} - \lambda_{1} \left( B_{{y}_{1}}+N_{y_{1}}\right)\right]; \  
E_{-|y_{1}}^{2}= \dfrac{1}{4} \left[ \mathbb{I} - \lambda_{1} \left( B_{{y}_{1}}- N_{y_{1}}\right)\right].\nonumber\\
\end{eqnarray}
	with $E_{+|y_{1}}=E_{+|y_{1}}^{1}+E_{+|y_{1}}^{2}$, $E_{-|y_{1}}=E_{-|y_{1}}^{1}+E_{-|y_{1}}^{2}$  and $ E_{+|y_{1}}+E_{-|y_{1}}=\mathbb{I}$. Note here that $M_{y_{1}}$ and $N_{y_{1}}$ are not unique, as explained in Appendix \ref{appg4}. However, such a non-uniqueness will not play any role as far as the optimal quantum value of the CHSH expression $\langle \mathcal{B}^{2} \rangle^{g=4}_Q$ is concerned. The $4$-POVMs given by Eq. (\ref{povm2q}) can be written in a general form as the sum of mutually commuting observables for any arbitrary dimensional system. One such generalized example for $g=4$ is given in the Appendix \ref{appg4}. 
	
If Bob$_1$'s instruments are characterized by four Kraus operators $K^{j}_{b_{1}|y_{1}}=\sqrt{E_{b_{1}|y_{1}}^{j}}$ with $j=1,2$,  then after Bob$_{1}$'s non-selective measurements the average reduced state shared by Alice and Bob$_{2}$ can be written as 
\begin{equation}\label{rhok}
\rho_{AB_{2}}^{g=4}=\frac{1}{2}\sum_{y_{1}=1}^{2} \sum_{b_{1}\in \pm}\sum_{j=1}^{2} \left(\mathbb{I} \otimes K^{j}_{b_{1}|y_{1}}\right)  \ \rho_{AB_{1}}^{g=4} \ \left(\mathbb{I} \otimes K^{j}_{b_{1}|y_{1}}\right)
\end{equation}
	
The Kraus operators are constructed as $K_{b_{1}|y_{1}}^{j}= \alpha_{4} \mathbb{I} + \beta_{4} X_{b_{1}|y_{1}}^{j} $ where $X_{+|y_{1}}^{1}= B_{y_{1}}+ M_{y_{1}}$, $X_{+|y_{1}}^{2}= B_{y_{1}}- M_{y_{1}}$, $X_{-|y_{1}}^{1}= -B_{y_{1}}- N_{y_{1}}$, $X_{-|y_{1}}^{2}= -B_{y_{1}}- N_{y_{1}}$ with $\sum_{b_{1},j}X_{b_{1}|y_{1}}^{j}=0$, where the parameters $\alpha_{4}$ and $\beta_{4}$ can be determined as 
\begin{eqnarray}\label{alpha4}
\alpha_4 = \dfrac{1}{4} \left[ \sqrt{\frac{1+3\lambda_{1}}{4}} + 3 \sqrt{\frac{1-\lambda_{1}}{4}}\right]\nonumber\\
\beta_4 = \dfrac{1}{4} \left[ \sqrt{\frac{1+3\lambda_{1}}{4}} - \sqrt{\frac{1-\lambda_{1}}{4}}\right]
\end{eqnarray}
The detailed analysis of how the values of $\alpha_{4}$ and $\beta_{4}$ are written in terms of $\lambda_{1}$ is crucial to this work which is placed in Appendix \ref{appg4}. Using Kraus operators given by Eq. (\ref{rhok}), the explicit form of $\rho_{AB_{2}}^{g=4}$ can be derived as
\begin{eqnarray}
\rho_{AB_{2}}^{g=4}	&=& 4 \alpha_{4}^{2} \ \rho_{AB_{1}}^{g=4} +2\beta_{4}^{2} \ \Bigg[\sum_{y_{1}=1}^{2} \qty(\mathbb{I} \otimes B_{y_{1}}) \ \rho_{AB_{1}}^{g=4} \ \qty(\mathbb{I} \otimes B_{y_{1}})\nonumber\\
&+& \sum_{{y_{1}}=1}^{2} \sum_{j=1}^{2}\qty(\mathbb{I} \otimes B_{y_{1}}^{j}) \ \rho_{AB_{1}}^{g=4} \ \qty(\mathbb{I} \otimes  B_{y_{1}}^{j}) \Bigg] .
\end{eqnarray}
	
Now, adopting the similar SOS approach as derived in Sec. \ref{sec3} for $g=2$ case, we evaluate the quantum value of the CHSH expression for Alice and Bob$_{2}$ as
\begin{equation}
\langle\mathcal{B}^2\rangle^{g=4}_Q= \Tr[\rho_{AB_{2}}^{g=4} \mathcal{B}]=4 (\alpha_4^2 + \beta_4^2 ) \ \langle \mathcal{B} \rangle_Q^{opt}
\end{equation}
Putting the values of $\alpha_{4}$ and $\beta_{4}$ from Eq. (\ref{alpha4}), we can write $\langle \mathcal{B}^{2} \rangle^{g=4}_Q$ in terms of only the unsharpness parameter as 
\begin{equation}\label{bg4}
\langle \mathcal{B}^{2} \rangle^{g=4}_Q=\dfrac{1}{4}\Big[\Big((3-\lambda_{1})+\sqrt{(1+3\lambda_{1})(1-\lambda_{1})}\Big)\Big]\langle \mathcal{B} \rangle_Q^{opt}
\end{equation}
Thus, to obtain advantage for Bob$_{2}$, two sequential Bob requires $\langle \mathcal{B}^{1} \rangle_Q^{g=4}, \langle \mathcal{B}^{2} \rangle_Q^{g=4}>2$. If Bob$_{2}$ performs sharp measurement (i.e., $\lambda_{2}=1$), then from Eq. (\ref{bg4}) we get the upper bound of $(\lambda_{1})_{\max} \approx 0.890$. 
	
Now, in order to compare the obtained CHSH value of Alice and Bob$_2$ corresponding to the $g=4$ case with that obtained in $g=2$ case, we recall the CHSH value for the latter case from Eq. (\ref{llbell1}) in terms of the unsharpness parameter 
\begin{equation}\label{llbell2}
\langle \mathcal{B}^{2} \rangle^{g=2}_Q=\dfrac{1}{2}\Big(1+\sqrt{1-\lambda_{1}^2}\Big)\langle \mathcal{B} \rangle_Q^{opt}
\end{equation}
	
Interestingly, from  Eqs. (\ref{bg4}) and (\ref{llbell2}), although it is evident that both the expressions given by $\langle \mathcal{B}^{2} \rangle^{g=4}_Q$  and $\langle \mathcal{B}^{2} \rangle^{g=2}_Q$ are optimized for the same state and observables, they are quantitatively inequivalent for any values of the unsharpness parameter within the allowed range. Such quantitative inequivalence between the empirical statistics of Alice and Bob$_2$ brings out the profound realization that the nature of the post-measurement reduced state of Alice and Bob$_1$ critically depends on the choice of degeneracy-breaking scheme.

\subsection{ Degeneracy-breaking measurement for $g=3$ case}\label{sec4b}
	
Now, for completeness, we examine the case when Bob$_{1}$ performs degeneracy-breaking measurement by considering $3$-POVMs. Intuitively it follows that $\langle \mathcal{B}^{2} \rangle^{g=2}_{Q}>\langle \mathcal{B}^{2} \rangle^{g=3}_{Q}>\langle\mathcal{B}^{2} \rangle^{g=4}_{Q}$. 
	
Here we construct the POVMs in $g=3$ case as $\{E_{+|y_{1}}^{12}\equiv E_{+|y_{1}}=(E_{+|y_{1}}^{1}+E_{+|y_{1}}^{2}), E_{-|y_{1}}^{1}$, $E_{-|y_{1}}^{2}\}$ or  \{$E_{+|y_{1}}^{1}$, $E_{+|y_{1}}^{2}, E_{-|y_{1}}^{12}\equiv E_{-|y_{1}}=(E_{-|y_{1}}^{1}+E_{-|y_{1}}^{2})$\}. Interestingly, $g=3$ case has only two symmetric permutation to implement the measurements where $E_{+|y_{1}}$ or $E_{-|y_{1}}$ provides the conventional POVMs corresponding to $g=2$. The Kraus operator for the first construction corresponding to the POVMs are given as $K_{+|y_{1}}^{12}= (\alpha_{2} \mathbb{I} +\beta_{2} B_{y_{1}})$, $K_{-|y_{1}}^{1}=(\alpha_{4} \mathbb{I}+\beta_{4} X_{-|y_{1}}^{1})$ and $K_{-|y_{1}}^{2}=(\alpha_{4} \mathbb{I}+\beta_{4} X_{-|y_{1}}^{2})$.
	Quantum value of the CHSH expression for Alice and Bob$_{1}$ $\langle \mathcal{B}^{1} \rangle^{g=3}_Q=\lambda_{1}\langle \mathcal{B} \rangle_Q^{opt}$ remains the same and irrespective of the value of $g$ .

	The quantum value of the CHSH expression for Alice and Bob$_{2}$ is calculated as 
	\begin{equation}
		\langle \mathcal{B}^{2} \rangle^{g=3}_Q=\Big[\alpha_{2} ^2+ 2(\alpha_{4}^2 + \beta_{4}^2 )\Big] \langle \mathcal{B} \rangle_Q^{opt}
	\end{equation} 
	Details of the derivation is given in Appendix \ref{appg3}. Writing $\langle \mathcal{B}^{2} \rangle^{g=3}_Q$ in terms $\lambda_{1}$, we get
\begin{eqnarray}\label{bg3}
\nonumber
\langle \mathcal{B}^{2} \rangle^{g=3}_Q=\dfrac{1}{8}\Big[(5-\lambda_{1})+ 2\sqrt{1-\lambda_{1}^2}+ \sqrt{\Big(1+3\lambda_{1} \Big)\Big( 1-\lambda_{1}\Big)}\Big] \langle \mathcal{B} \rangle_Q^{opt}\\
\end{eqnarray}
If Bob$_{2}$ performs sharp measurement (i.e., $\lambda_{2}=1$), then from Eq. (\ref{bg3}) we get the upper bound of $(\lambda_{1})_{\max} \approx 0.865$. It is seen that $\langle \mathcal{B}^{2} \rangle^{g=3}_Q$ is also optimized for the same state, observables for which $\langle \mathcal{B}^{2} \rangle^{g=2}_Q$ is optimized. This is due to the fact that $\langle \mathcal{B} \rangle_Q^{opt}$ is the same for both the Eqs. (\ref{llbell2}) and (\ref{bg3}) and coefficients are only dependent on$\lambda_{1}$. 

\subsection{$g$-POVM certification argument}\label{sec4c}

\begin{figure}[ht]\centering 
{{\includegraphics[width=0.9\linewidth]{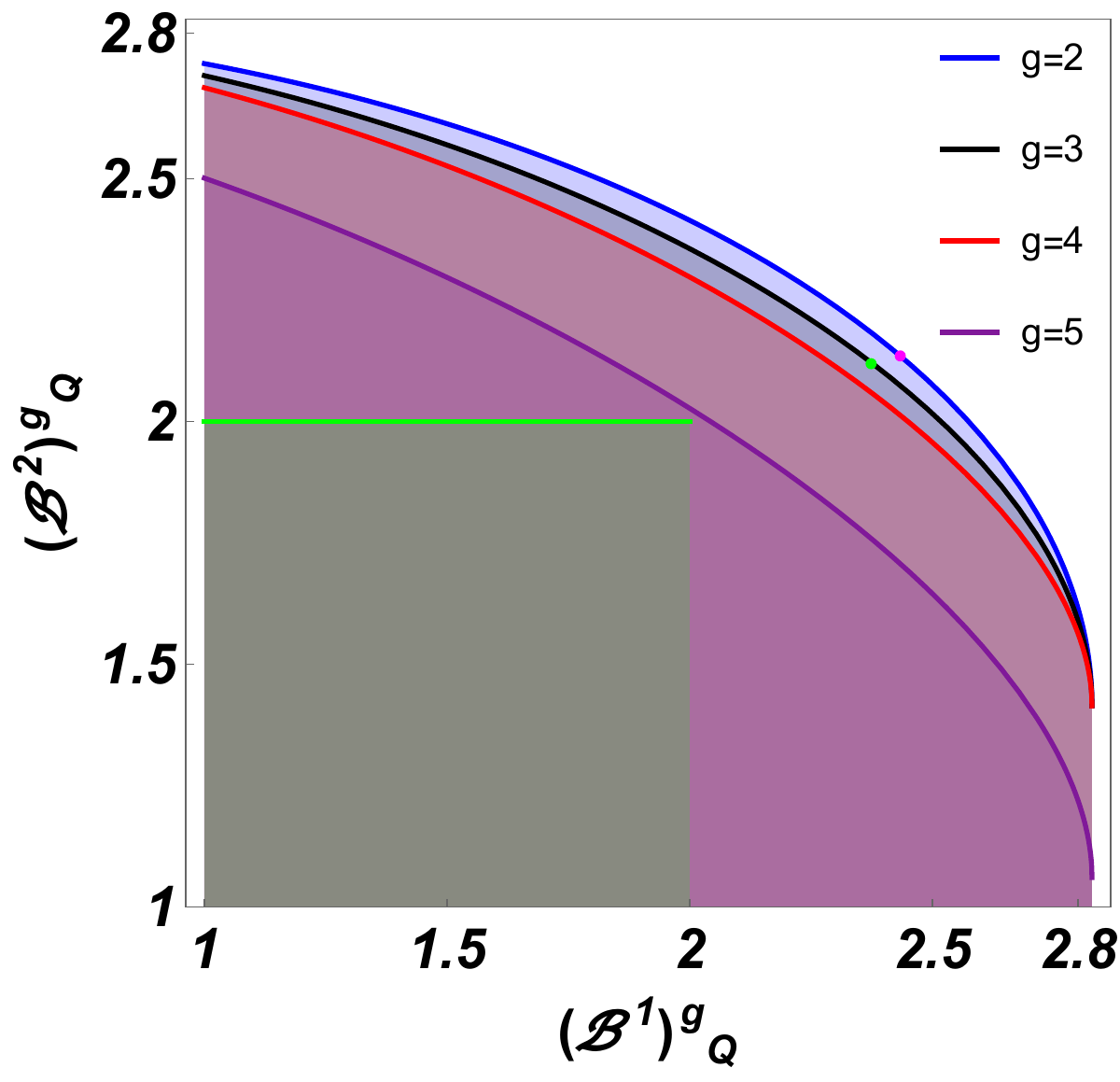}}}
\caption{\footnotesize \centering Optimal trade-off between the quantum value of the CHSH inequality for Alice-Bob$_1$ and Alice-Bob$_2$ for different $g$-POVMs is shown by the solid curves with the shaded portion giving the suboptimal range. The solid green line is for the classical bound of the CHSH inequality for the same two observers. }
\label{lamb}
\end{figure}

{We note that $\langle \mathcal{B}^{2} \rangle^{g=2}_Q>\langle \mathcal{B}^{2} \rangle^{g=3}_Q >\langle \mathcal{B}^{2} \rangle^{g=4}_Q$ for any nonzero values of $\lambda_{1}$, we can always certify the number of POVMs used by Bob$_{1}$ only from the observed statistics. This, therefore, enables the DI certification of degeneracy-breaking measurements in the sequential CHSH scenario. 

We start by noting again that $\langle \mathcal{B}^{1}\rangle_Q=\lambda_{1}\langle \mathcal{B}\rangle_Q^{opt}$ is independent of the value of $g$. By using Eq. (\ref{llbell2}), we can write $\langle \mathcal{B}^{2}\rangle_Q^{g=2}$ as a function of $\langle \mathcal{B}^{1}\rangle_Q$, so that
\begin{eqnarray}
\label{td1}
\langle \mathcal{B}^{2} \rangle^{g=2}_Q&=&\dfrac{1}{2}\qty(\langle\mathcal{B} \rangle_Q^{opt}+\sqrt{\qty(\langle \mathcal{B}\rangle_{Q}^{opt})^{2}-\Big(\langle \mathcal{B}^{1}\rangle_{Q}\Big)^{2}})
\end{eqnarray}
Since $\langle\mathcal{B} \rangle_Q^{opt}=2\sqrt{2}$, then $\langle \mathcal{B}^{2} \rangle^{g=2}_Q$ is the function of $\langle \mathcal{B}^{1} \rangle_Q$ only. The Eq. (\ref{td1}) represents the trade-off between the quantum values of CHSH expression for Alice-Bob$_{1}$ ($\langle \mathcal{B}^{1} \rangle_Q$) and Alice-Bob$_{2}$ ($\langle \mathcal{B}^{2} \rangle^{g=2}_Q$) which is plotted in FIG. \ref{lamb}. The sub-optimal quantum values $\langle \mathcal{B}^{1} \rangle_Q$ and $\langle \mathcal{B}^{2} \rangle^{g=2}_Q$ form an optimal pair $\{\langle \mathcal{B}^{1} \rangle_Q,\langle \mathcal{B}^{2} \rangle^{g=2}_Q\}$ of sequential quantum violation of CHSH inequality.}

{Similarly, using Eqs. (\ref{bg4}) an (\ref{bg3})  one finds that  $\langle \mathcal{B}^{2}\rangle_Q^{g=4}$ and $\langle \mathcal{B}^{2}\rangle_Q^{g=3}$ can also be written as a sole function of $\langle \mathcal{B}^{1}\rangle_Q$. This is depicted in FIG. \ref{lamb} for different values of $g\leq5$. The explicit calculation for $g=5$ is placed in Appendix \ref{appg}. It is crucial to note that the trade-off curve for each $g$ is nonoverlapping. Importantly, every optimal pair $\{\langle \mathcal{B}^{1} \rangle_Q,\langle \mathcal{B}^{2} \rangle^{g}_Q\}$ corresponding to different non-overlapping curves uniquely certify each upper bound on $g$-POVMs that Bob$_{1}$ may have implemented in a particular run of the experiment. 

Let us now discuss certification of $g$-POVMs by considering the trade-off relations corresponding to the optimal pair  $\{\langle \mathcal{B}^{1}\rangle_Q,\langle\mathcal{B}^{2}\rangle_Q^{g}\}$. As an example, experimentally one gets the quantum values of CHSH expressions for Alice-Bob$_{2}$ as  $\langle \mathcal{B}^{2}\rangle_Q^{g}=2.12$. This value can come from two different $g$-POVMs measurements. But the experimentalist also gets the quantum values of CHSH expressions for  Alice-Bob$_{1}$ as $\langle \mathcal{B}^{1}\rangle_Q=2.37$.  Then, the pair $\{\langle\mathcal{B}^{1}\rangle_Q,\langle\mathcal{B}^{2}\rangle_Q^{g}\}\approx\{2.37,2.12\}$ represents a unique point which is marked as a green dot in FIG. \ref{lamb} that lies on the curve for $g=3$. This then ensures that Bob$_{1}$ must have used the value of $g\leq 3$.

Similarly, let in another run of experiment one realizes the optimal pair  $\{\langle\mathcal{B}^{1}\rangle_Q,\langle\mathcal{B}^{2}\rangle_Q^{g}\}\approx\{2.43,2.13\}$ which is marked as a magenta dot in between the curves for $g=2$ and $g=3$.  This simply ensures that Bob$_{1}$ must not have performed degeneracy-breaking measurement ($g>2$) and he should have performed degeneracy-preserving $g=2$ measurement, but there is noise in the measurement. Hence, from the quantum value of the optimal pair  $\{\langle\mathcal{B}^{1}\rangle_Q,\langle\mathcal{B}^{2}\rangle_Q^{g}\}$, the upper bound on the number of $g$-POVMs used by Bob$_{1}$ can be certified.}

\section{Certification of $g$-POVMs using certified $\lambda_{1}$}\label{sec5}

\begin{figure}
\centering
\subfloat[\centering The violation $\langle \mathcal{B}^{2} \rangle^{g}_Q$ is plotted against $\lambda_{1}$ for $g=2,3,4$ and $5$  ]{{\includegraphics[width=8.3cm]{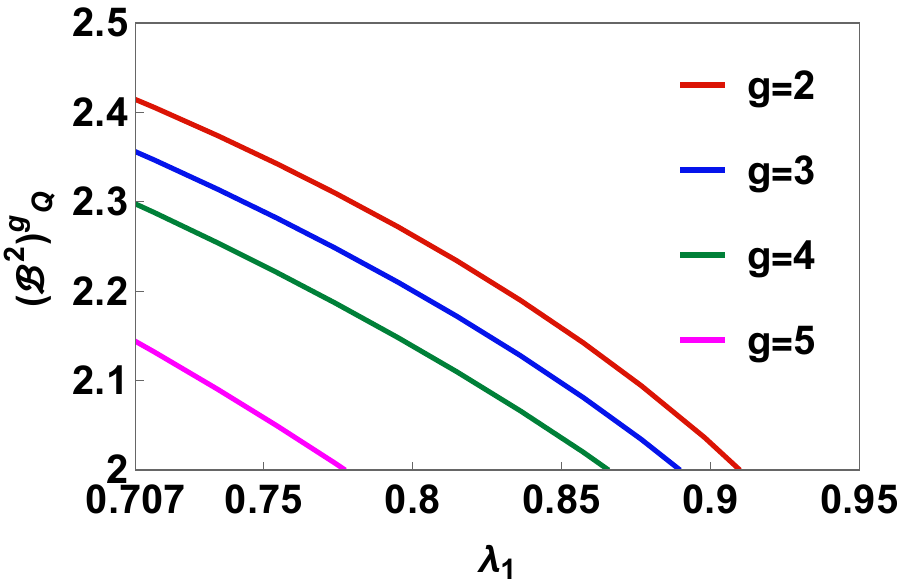} }}
\qquad
\subfloat[\centering Points showing the quantum value of the CHSH expression $\langle \mathcal{B}^{2} \rangle^{g}_Q$ of Bob$_{2}$ for distinct values of $\lambda_{1}$ and g-POVMs ]{{\includegraphics[width=8cm]{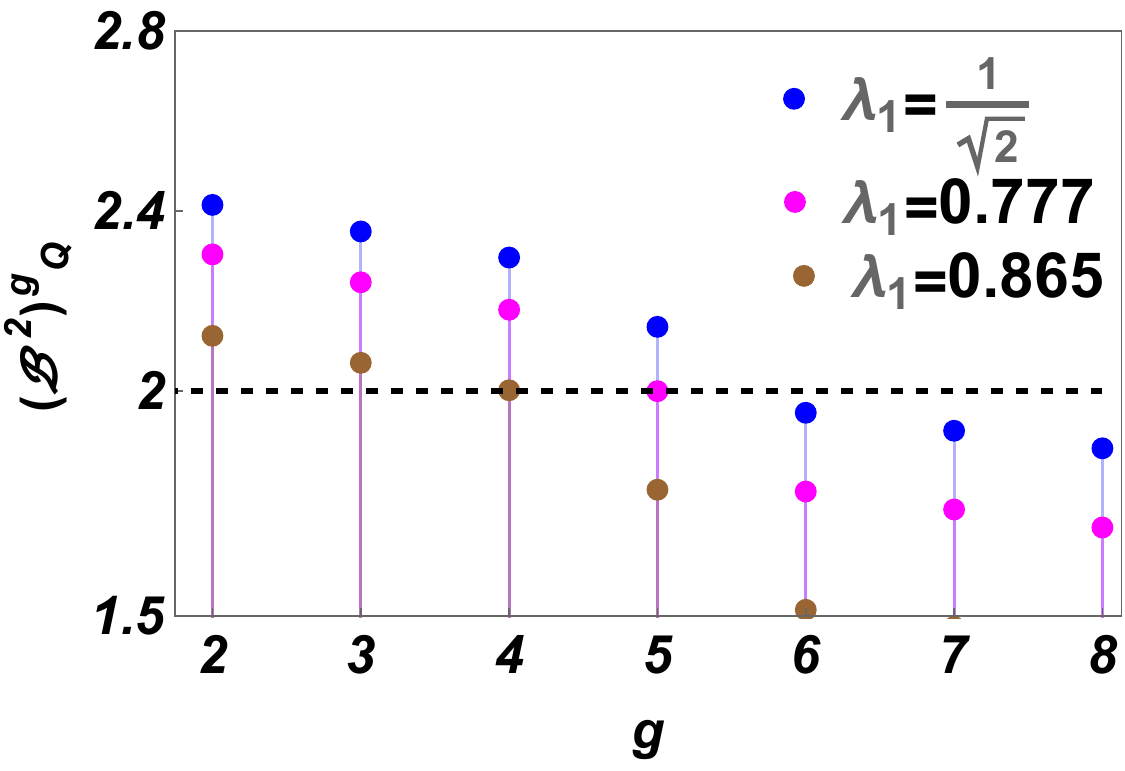} }}
\caption{Graphs showing the dependence of $\langle \mathcal{B}^{2} \rangle^{g}_Q$ on the values of $g$ ( $\lambda_{1}$) of implemented measurement  of Bob$_{1}$ for particular values of $\lambda_{1}$ ( $g$ ). This captures the fact that the more the Bob$_{1}$ breaks the degeneracy of the measuring system the less should be the critical value of $\lambda_{1}$ that provides the quantum advantage to Both the observers. }  
\label{join}
\end{figure}

	{To add more regarding the $g$-POVMs certification we may start with a certified unsharpness parameter which can be done by additional measurements.  If the measurement is implemented with instruments having unsharpness parameter greater than the critical value $\lambda_{1}^{\ast}=1/\sqrt{2}$, the optimal quantum values of Bell expressions for $g=2,3$ and $4$ satisfy $\langle \mathcal{B}^{2} \rangle^{g=2}_Q>\langle \mathcal{B}^{2} \rangle^{g=3}_Q>\langle \mathcal{B}^{2} \rangle^{g=4}_Q$ as depicted in FIG. \ref{join}a which captures the quantitative analysis of the different degrees of degeneracy-breaking measurements. This is explicitly discussed in the following section.}

	It is already clear that both the unsharpness as well as the g-POVM of the implemented measurement result in the loss of shared quantum correlation between Alice and Bob$_{2}$.  However, for the case of $g=2$, there is no additional loss of correlation due to degeneracy-breaking. Then for $g=2$, the only viable parameter that puts restriction on the simultaneous CHSH violation due to the loss of correlation is the unsharpness parameter $\lambda_{1}$ of Bob$_1$'s measurement. Consequently, the more the value of the unsharpness parameter, the less correlation remains in the relayed system. This then directly follows that for $g=2$ the CHSH value $\langle \mathcal{B}^{2} \rangle^{g}_Q$ of Bob$_{2}$ is a monotonically decreasing function of $\lambda_{1}$ as depicted by the red curve in FIG.\ref{join}a. 
 
  Now, to ensure a simultaneous violation of the CHSH inequality by both the observer, the residual correlation of the system after Bob$_{1}$'s measurement must be greater than a certain threshold value. While the lower bound on the unsharpness parameter $\lambda_1$ is fixed  when Bob$_1$ just starts getting CHSH violation, the requirement of a simultaneous CHSH violation for both the observer fixes upper bound of $\lambda_1$. Thus, in the scenario of $g$-POVM measurement with $g=2$, it is possible to certify a range of unsharpness parameter only from the input-output statistics. Such range of the unsharpness parameter can be evaluated from Eq. \eqref{llbell2} as $0.707<\lambda_{1}\leq0.912$. At this point, an obvious question may arise what happens when one considers the scenario beyond $g=2$ case? Is it possible for both the observer to get simultaneous quantum violation of CHSH inequality for the case of  $g>2$? If yes, what are the upper bounds on $\lambda_1$ for different $g(>2)$ values that constitute such sustainable quantum violations? This is explicitly discussed in the following.

Let us first discuss the possibility of having simultaneous quantum violations of CHSH inequalities by both observers corresponding to different $g$-POVM $(g>2)$ measurements for an arbitrary dimensional quantum system. Obviously, as $g$-value increases, the system is subjected to more disturbance and vice versa. It is then evident that if more correlation is lost due to a certain degeneracy-breaking interaction, one has to implement more unsharp ( less $\lambda_1$) measurement to compensate for that loss. Then, from a comparative point of view, it is clear that although for a specific $g$-value $\langle \mathcal{B}^{2} \rangle^{g}_Q$ remains a monotonically decreasing function of $\lambda_1$, for a greater value of $g$ the upper bound to the unsharpness parameter is lesser as shown in FIG.\ref{join}a. The upper bounds to the unsharpness parameter $\lambda_{1}$ for different $g$-POVMs measurements are evaluated explicitly in Appendix B. In particular, for the case of $g=3,g=4$, $g=5$ and $g=6$ the threshold values are $\lambda_{1}=0.890$, $\lambda_{1}=0.865$, $\lambda_{1}=0.777$ and $\lambda_{1}=0.686$, respectively. It then follows that for any $g>5$ measurement, the threshold value that gives the quantum violation of the CHSH inequality to Bob$_{2}$ is insufficient for Bob$_{1}$ to get any violation.  Thus, for $g>5$, there exist no values of $\lambda_{1}$ for which both the independent observers get the simultaneous quantum violations of the CHSH inequality. However, such simultaneous violation persists as long as the implemented measurements of Bob$_{1}$ correspond to $g\leq 5$. We argue that such quantum violations of the CHSH inequality by more than one sequential Bob can be used as a resource to certify the g-POVMs of the implemented measurement.
	
	In order exemplify the discussed facts, let us now contemplate the plot depicted in FIG.\ref{join}b, which shows the dependence of $\langle \mathcal{B}^{2} \rangle^{g}_Q$ on $g$ for particular values of $\lambda_{1}$. In FIG.\ref{join}b it is explicitly shown that for the above-mentioned threshold values $\lambda_{1}=0.865$ and $\lambda_{1}=0.777$,  the points that reside exactly on the dotted line for $g=4$ and $g=5$ respectively, correspond to the CHSH value $\langle \mathcal{B}^{2} \rangle^{g}_Q=2$. It also shows that for $g=6$ even if the system is disturbed minimally ($\lambda_{1}=1/\sqrt{2}$), the violation of the CHSH inequality is forbidden for Bob$_{2}$.

	\section{Summary and discussions}\label{sec6}

	Our work may pave the path for a number of future studies. First, in this work, we confined ourselves to the simplest 2-2-2 Bell scenario in which it is not possible to certify the degeneracy breaking measurement beyond $g=5$. Thus one may expect whether it is possible to certify more degree of the degeneracy-breaking measurement ($g>6$) by going beyond the 2-2-2 scenario. For this purpose, to begin with, one may proceed by going beyond the 2-measurement setting scenario and invoke the recently proposed \cite{pan19} family of Bell inequalities which was introduced as a dimension witness \cite{pan2020}. Such a particular family of Bell inequalities has a very unique power that shows increasing optimal quantum value with an increase in the number of measurement settings than that of existing the CHSH or the Gisin's elegant Bell inequality \cite{gisin1}. Moreover, by invoking higher outcome Bell inequalities like the CGLMP inequality \cite{gisin2} for the 3-outcome scenario, one can also revisit the degeneracy breaking scheme. 
	
	Second, since degeneracy-breaking measurement can be certified in the DI way, it will be interesting if one uses it as a resource for DI information processing tasks such as the generation of more randomness in secure communication tasks or in the generation of secure key for cryptographic applications. Another direction may be explored going beyond the nonlocal scenario. Since nonlocality is the strongest form of quantum correlation, one can invoke the steering or entanglement scenario in order to certify the degeneracy-breaking measurement for more than $g=5$. Those could be an interesting avenue for future research.

	\section*{Acknowledgments}
We thank Sauradeep Sasmal for raising critical comments and for helping us to finalize this work. PR, SM, and AKP acknowledge the support from the project DST/ICPS/QuST/Theme 1/2019/4. S.S.M. acknowledges the UGC fellowship [Fellowship No.16-9(June 2018)/2019(NET/CSIR)].
	\begin{widetext}
		\appendix
		\section{Sum-of-square (SOS) approach for the derivation of optimal quantum violation without specifying the dimension of the system }
		\label{appsos}
	 Here we provide a derivation of the $\langle \mathcal{B} \rangle^{opt}_{Q}$ without assuming  the Hilbert space dimension by introducing an elegant sum-of-squares (SOS) \cite{sneha} approach. One can argue that there is a positive semi-definite operator $\eta \geq 0$, that can be expressed as 
  \begin{eqnarray}
  (\eta)_{Q}=\Delta -\langle \mathcal{B} \rangle_{Q} 
 \end{eqnarray}
 where  $\Delta\geq 0$. This can be proven by considering two suitable positive operators, $M_{1}$ and $M_{2}$, which are polynomial functions of   $A_{x}$ and $B_y$, so that 

\begin{align}
	\label {g1}
	\eta =\frac{1}{2} \left( \omega_{1} M_1^\dagger M_1+\omega_{2} M_2^\dagger M_2\right). 
\end{align}
For our purpose, we suitably choose $ M_1 $ and $M_{2}$ as  

\begin{align}
	\label{mi}
	M_{1}|\psi\rangle_{AB}=\frac{A_{1}+A_{2}}{\omega_{1}} |\psi\rangle_{AB} -B_{1} |\psi\rangle_{AB}\\
	\nonumber
	M_{2}|\psi\rangle_{AB}=\frac{A_{1}-A_{2}}{\omega_{1}} |\psi\rangle_{AB} -B_{2} |\psi\rangle_{AB}
\end{align} 
where $\omega_{1} =||\left(A_{1}+A_{2}\right)|\psi\rangle_{AB}||_{2}$ and $\omega_{2} =||\left(A_{1}-A_{2}\right)|\psi\rangle_{AB}||_{2}$ for a quantum state $|\psi\rangle_{AB}$. Here $||.||_{2}$ is the Frobenius norm of a vector, $||\ \mathcal{O}\ ||_2=\sqrt{Tr[\mathcal{O}^{\dagger}\mathcal{O} \rho]}$. Plugging Eq. (\ref{mi}) into Eq. (\ref{g1}) and using that $A_{x}^{\dagger} A_{x}=B_y^{\dagger} B_y=\mathbb{I} $, we obtain
\begin{align}
	\langle \mathcal{B} \rangle_{Q}= \left(\omega_{1}+\omega_{2}\right) -(\eta)_{Q} .
\end{align}
It then follows that the optimal value of $\langle \mathcal{B} \rangle_{Q}$ can be obtained when $(\eta)_{Q}= 0$, therefore,   

\begin{eqnarray}\label{optbnn}
\langle \mathcal{B} \rangle_{Q}^{opt} &=&\max\left(\omega_{1}+\omega_{2}\right)\nonumber\\
&=&\underset{{}}{\max}\left(\sqrt{2+\langle \{A_1, A_2\}\rangle}+\sqrt{2-\langle \{A_1, A_2\}\rangle}\right) .
\end{eqnarray}
Thus, the maximization requires $ \{A_1, A_2\}=0$ implying Alice's observables have to be anticommuting. In turn, the optimization condition provides
\begin{equation}
\Tr[\left(A_{1}+A_{2}\right) \ |\psi\rangle_{AB}\bra{\psi}]=\Tr[\left(A_{1}-A_{2}\right) \ |\psi\rangle_{AB}\bra{\psi}]
\end{equation}
which gives $\omega_{1}=\omega_{2}=\sqrt{2}$, and consequently the optimal value $\langle \mathcal{B} \rangle_{Q}^{opt}=2\sqrt{2}$. 

The explicit conditions for the optimization are 
\begin{eqnarray}\label{l12}
M_{1}|\psi\rangle_{AB}=0 \implies \ B_{1}=\frac{A_{1}+A_{2}}{\sqrt{2}} \ \ ; \ \ \text{and} \ \ M_{2}|\psi\rangle_{AB}=0 \implies \ B_{2}=\frac{A_{1}-A_{2}}{\sqrt{2}}.
\end{eqnarray} 
The above Eq. (\ref{l12}) follows that $\{B_1, B_2\}=0$, i.e., Bob's observables are also anticommuting. Moreover, for the state $\rho_{AB}=\ket{\psi_{AB}}\bra{\psi_{AB}} \in \mathcal{H}_{A}^{d} \otimes \mathcal{H}_{B}^d $, the optimal violation is obtained when $\Tr[\qty(B_{1}\otimes B_{1}) \ \rho_{AB}]=\Tr[\qty(B_{2}\otimes B_{2}) \ \rho_{AB}]=1$. It can be shown that is possible only if $\ket{\psi}_{AB}$ is a maximally entangled state. Thus, the optimal quantum value $\langle \mathcal{B} \rangle_{Q}^{opt}$ uniquely certifies the state and observables. 

Note that, here we have derived the quantum optimal value of the CHSH expression without specifying the dimension of the system.

{\section{Detail derivation of Eqs.~(\ref{primeobs}) and (\ref{maxchsh}) in the main text}\label{appsos2}

The quantum value of the CHSH expression for Alice and Bob$_{2}$ in the second line of Eq. (\ref{b2bell}) in the main text is derived as follows. From the first line of Eq. (\ref{b2bell}), we have $\langle \mathcal{B}^{2} \rangle_Q^{g=2}= \underset{\rho_{AB_{1}},\{A_{x}\},\{B_{y}\}}{\max}\qty(\Tr[\rho_{AB_{2}} \mathcal{B}])$. By putting the $\rho_{AB_{2}}$ and $\mathcal{B}$ from Eq. (\ref{lu2}) and Eq. (\ref{bellor}) respectively, we get
 \begin{eqnarray}\label{b2bellapp}
\langle \mathcal{B}^{2} \rangle_Q^{g=2}&=&\underset{\rho_{AB_{1}},\{A_{x}\},\{B_{y}\}}{\max}\qty(\Tr[\qty(2\alpha_{2}^{2}\rho_{AB_{1}}+\beta_{2}^{2}\qty(B_{1}\rho_{AB_{1}}B_{1}+B_{2}\rho_{AB_{1}}B_{2})) \qty[\qty(A_{1}+A_{2}) B_{1} +\qty(A_{1}-A_{2}) B_{2}]])\nonumber  \  \ \text{[from Eqs. (2) and (3)]}\\
&=&\underset{\rho_{AB_{1}},\{A_{x}\},\{B_{y}\}}{\max}\qty( \Tr[\qty(2\alpha_{2}^{2}\qty[\qty(A_{1}+A_{2}) B_{1} +\qty(A_{1}-A_{2}) B_{2}]+\beta_{2}^{2} \qty[\qty(A_{1}+A_{2})\qty(B_{1}+B_{2}B_{1}B_{2})+\qty(A_{1}-A_{2})\qty(B_{2}+B_{1}B_{2}B_{1})]) \ \rho_{AB_{1}}])\nonumber\\
&=&\underset{\rho_{AB_{1}},\{A_{x}\},\{B_{y}\}}{\max}\qty( \Tr[\qty(\qty(A_{1}+A_{2})\qty[2\alpha_{2}^{2}B_{1}+\beta_{2}^{2}\qty(B_{1}+B_{2}B_{1}B_{2})]+ \qty(A_{1}-A_{2})\qty[2\alpha_{2}^{2}B_{2}+\beta_{2}^{2}\qty(B_{1}B_{2}B_{1}+B_{2})]) \ \rho_{AB_{1}}])\nonumber\\
&=&\underset{\rho_{AB_{1}},\{A_{x}\},\{B_{y}\}}{\max}\qty(\Tr[\qty(\qty(A_{1}+A_{2})\qty[\qty(2\alpha_{2}^{2}+\beta_{2}^{2})B_{1}+\beta_{2}^{2} \ B_{2}B_{1}B_{2}]+ \qty(A_{1}-A_{2})\qty[\qty(2\alpha_{2}^{2}+\beta_{2}^{2})B_{2}+\beta_{2}^{2} \ B_{1}B_{2}B_{1}]) \ \rho_{AB_{1}}])\nonumber\\
&=&\underset{\rho_{AB_{1}},\{A_{x}\},\{B_{y}\}}{\max}\qty(\Tr[\rho_{AB_{1}}\qty((A_{1}+A_{2})B_{1}^{\prime}+(A_{1}-A_{2})B_{2}^{\prime})]),
\end{eqnarray}
where 
\begin{eqnarray}
    B_{1}^{\prime}&=&\qty(2\alpha_{2}^{2}+\beta_{2}^{2})B_{1}+\beta_{2}^{2} \ B_{2}B_{1}B_{2}\nonumber \\
    B_{2}^{\prime}&=&\qty(2\alpha_{2}^{2}+\beta_{2}^{2})B_{2}+\beta_{2}^{2} \ B_{1}B_{2}B_{1}
\end{eqnarray}
which is the Eq. (\ref{primeobs}) in the main text. It is interesting to note that Eq. (\ref{b2bell}) has the similar form of CHSH expression with the exception that $B_{1}^{\prime}$ and $B_{2}^{\prime}$ are unnormalized. To derive the maximum quantum value of $\langle \mathcal{B}^{2} \rangle_{Q}^{g=2}$ we do not assume the Hilbert space dimension. An elegant SOS approach similar to the one provided in Appendix \ref{appsos} enables such a derivation as given below.\\

For this we assume a positive semi-definite operator $\eta^{\prime} \geq 0$, that can be expressed as 
  \begin{eqnarray}
  (\eta^{\prime})_{Q}=\Delta -\langle \mathcal{B}^{2} \rangle_{Q}^{g=2}
 \end{eqnarray}
Clearly, when $ (\eta^{\prime})_{Q}=0$, we get the maximum value of $\langle \mathcal{B}^{2} \rangle_{Q}^{g=2}$. This can be proven by considering two suitable positive operators, $M_{1}$ and $M_{2}$, which are polynomial functions of   $A_{x}$ and $B_y$, so that 

\begin{align}
	\label {g1}
	\eta^{\prime} =\frac{1}{2} \left( \omega_{1}\omega_{1}^{\prime} M_1^\dagger M_1+\omega_{2} \omega_{2}^{\prime} M_2^\dagger M_2\right). 
\end{align}
where $\omega_{i},\omega_{i}^{\prime} \ \forall i \in \{1,2\}$ are positive numbers. For our purpose, we suitably choose $ M_1 $ and $M_{2}$ as  

\begin{align}
	\label{mi}
	M_{1}|\psi\rangle_{AB}=\frac{A_{1}+A_{2}}{\omega_{1}} |\psi\rangle_{AB} -\dfrac{B_{1}^{\prime}}{\omega_{1}^{\prime}}|\psi\rangle_{AB}\\
	\nonumber
	M_{2}|\psi\rangle_{AB}=\frac{A_{1}-A_{2}}{\omega_{2}}|\psi\rangle_{AB} -\dfrac{B_{2}^{\prime}}{\omega_{2}^{\prime}}|\psi\rangle_{AB}
\end{align}
where $\omega_{1} =||\left(A_{1}+A_{2}\right)|\psi\rangle_{AB}||_{2}$, $\omega_{1}^{\prime} =|| B_{1}^{\prime}|\psi\rangle_{AB} ||_{2}$, $\omega_{2} =||\left(A_{1}-A_{2}\right)|\psi\rangle_{AB}||_{2}$ and $\omega_{2}^{\prime} =|| B_{2}^{\prime}|\psi\rangle_{AB}||_{2}$. Here $||.||_{2}$ is the Frobenius norm of a vector, $||\ \mathcal{O}|\psi\rangle\ ||_2=\sqrt{Tr[\mathcal{O}^{\dagger}\mathcal{O} \rho]}$. Plugging Eq. (\ref{mi}) into Eq. (\ref{g1}) and using that $A_{x}^{\dagger} A_{x}=B_y^{\dagger} B_y=\mathbb{I} $, we obtain
\begin{eqnarray}\label{sosforwp}
(\eta)_{Q} &=& \qty[\frac{\omega_{1}^{\prime}}{2\omega_{1}}\qty(A_{1}+A_{2})^{2}+ \frac{\omega_{2}^{\prime}}{2\omega_{2}}\qty(A_{1}-A_{2})^{2}+\frac{\omega_{1}}{2\omega_{1}^{\prime}}\qty(B_{1}^{\prime})^{2}+\frac{\omega_{2}}{2\omega_{2}^{\prime}}\qty(B_{2}^{\prime})^{2}] - \langle \mathcal{B}^{2} \rangle_{Q}^{g=2}
\end{eqnarray}
Further simplification straightforwardly provides  
\begin{eqnarray}
\langle \mathcal{B}^{2} \rangle_{Q}^{g=2}&=& \left(\omega_{1}\omega_{1}^{\prime}+\omega_{2}\omega_{2}^{\prime}\right) -(\eta^{\prime})_{Q} .
\end{eqnarray}

It then follows that the optimal value of $\langle \mathcal{B}^{2} \rangle_{Q}^{g=2}$ can be obtained when $(\eta^{\prime})_{Q}= 0$, therefore,   

\begin{eqnarray}\label{optbnnp}
\langle \mathcal{B}^{2} \rangle_{Q}^{g=2} &=&{\max}\qty(\omega_{1}\omega_{1}^{\prime}+\omega_{2}\omega_{2}^{\prime})
\end{eqnarray}\\
which is expressed in Eq. (\ref{maxchsh}) in the main text.}

\section{Proof to show Bob$_1$ and Bob$_2$ self-test same set of observables}	\label{appobser}

The standard sequential scenario comprises of one Alice who performs sharp measurement, and an two independent Bobs (Bob$_{1}$ and Bob$_{2}$) who perform unsharp measurements with respective unsharpness parameters $\lambda_{k}\ \forall k \in \{1,2\}$ sequentially. Let, Bob$_{1}$ performs measurement of $B_{1}$ and $B_{2}$ on his local subsystems upon receiving input $y_{1}\in \{1,2\}$ on the entangled state $\rho_{AB_{1}}$. After Bob$_{1}$'s measurement, he relays his subsystem to Bob$_{2}$. Upon receiving input $y_{2}\in \{1,2\}$, Bob$_{2}$ performs measurements of the observables $B_{3}$ and $B_{4}$ on the state $\rho_{AB_{1}}$ with unsharpness parameter $\lambda_{2}$, producing outputs $B_{2}\in \{0,1\}$.  The CHSH expression between Alice and Bob$_{2}$ is given by
\begin{equation}\label{bellapp}
\mathcal{B}^{2}=\left(A_{1}+A_{2}\right) B_{3} +\left(A_{1}-A_{2}\right) B_{4}.
\end{equation}

After the measurement of Bob$_{1}$, the reduced state averaged over Bob$_{1}$'s measurements and outcomes using Eqs. (\ref{kraus}) and (\ref{albe}) from the main text is given by
\begin{eqnarray}\label{lu2app}
\rho_{AB_{2}}&=&\dfrac{1}{2} \sum_{b_1 \in \{ +,-\}}\sum_{y_{1}=1}^{2} \left(\mathbb{I} \otimes K_{b_{1}|y_{1}}\right)  \rho_{AB_{1}}  \left(\mathbb{I} \otimes K_{b_{1}|y_{1}}\right)=\dfrac{1}{2} \Big[ 4\alpha_2^2 \  \rho_{AB_{1}} +2 \beta_2^2 \sum_{y_{1}=1}^{2} (\mathbb{I} \otimes B_{y_{1}})\rho_{AB_{1}}(\mathbb{I} \otimes B_{y_{1}}) \Big].
\end{eqnarray}

Now, putting $\rho_{AB_{2}}$ from Eq. (6) in the main text, the above CHSH expression is reduced to the following
\begin{eqnarray}\label{a1}	\langle\mathcal{B}^{2}\rangle_Q^{g=2}&=&\Tr[\rho_{AB_{2}}\mathcal{B}^{2}]=\Tr\Big[\rho_{AB_{1}}\left(\Big(A_{1}+A_{2}\Big) \bar{B}+\Big((A_{1}-A_{2})\bar{\bar{B}}\right)\Big]
\end{eqnarray}
where 
\begin{eqnarray}
\bar{B}&=&2\alpha^2_{2}B_{3}+\beta^2_{2}\Big(B_{1}B_{3}B_{1}+B_{2}B_{3}B_{2}\Big) \nonumber \\
\bar{\bar{B}}&=&2\alpha^2_{2}B_{4}+\beta^2_{2}\Big(B_{1}B_{4}B_{1}+B_{2}B_{4}B_{2}\Big),
\end{eqnarray}
with $\alpha_{2}=\frac{1}{2}\left[\sqrt{\frac{(1+\lambda_{1})}{2}}+ \sqrt{\frac{(1-\lambda_{1})}{2}}\ \right]$ and $\beta_{2}= \frac{1}{2}\left[\sqrt{\frac{(1+\lambda_{1})}{2}}-\sqrt{\frac{(1-\lambda_{1})}{2}}\ \right]$. Also,  $ \alpha^{2}_{2}=\frac{1}{4}\Big(1+a\Big)$ and $\beta^{2}_{2}=\frac{1}{4}\Big(1-a\Big)$ where $ a=\sqrt{1-\lambda^{2}}\geq 0$.
   
Putting the values of $\alpha_{2}^{2}$ and $\beta_{2}^{2}$, we can re-write $\bar{B}$ and $\bar{\bar{B}}$ as
\begin{equation}
    \bar{B}=\frac{1+a}{2}B_{3}+\frac{1-a}{4}\Big(B_{1}B_{3}B_{1}+B_{2}B_{3}B_{2}\Big); \ \ \ \ 
    \bar{\bar{B}}=\frac{1+a}{2}B_{4}+\frac{1-a}{4}\Big(B_{1}B_{4}B_{1}+B_{2}B_{4}B_{2}\Big).
\end{equation}

The optimal quantum value of $\langle\mathcal{B}^{2}\rangle_Q^{g=2}$ requires Bob's observables to be mutually anti-commuting, as proved in the manuscript. Hence, for Bob's (unnormalized) observables, we require,
\begin{eqnarray} \label{ac}
    &&\{\bar{B},\bar{\bar{B}}\}=0\nonumber \\
   &&\Rightarrow\frac{(1+a)^{2}}{4}\{B_{3},B_{4}\}+\frac{(1-a^{2})}{8}\{B_{3},(B_{1}B_{4}B_{1}+B_{2}B_{4}B_{2})\}\nonumber \\
   &&+\frac{(1-a^{2})}{8}\{B_{4},(B_{1}B_{3}B_{1}+B_{2}B_{3}B_{2})\}+\frac{(1-a)^{2}}{16}\{(B_{1}B_{3}B_{1}+B_{2}B_{3}B_{2}),(B_{1}B_{4}B_{1}+B_{2}B_{4}B_{2})\}=0
\end{eqnarray}

Now, rearranging the above Eq. (\ref{ac}) in terms of different degree of the co-efficient $a$, we will obtain three terms, each of which independently requires to be zero following the anti-commutativity relation. Thus, we obtain the  following three conditions as

 \begin{eqnarray} \label{c1}  &&4\{B_{3},B_{4}\}+2\{B_{3},(B_{1}B_{4}B_{1}+B_{2}B_{4}B_{2})\}+2\{B_{4},(B_{1}B_{3}B_{1}+B_{2}B_{3}B_{2})\}
 \\
 \nonumber
     &+&\{(B_{1}B_{3}B_{1}+B_{2}B_{3}B_{2}),(B_{1}B_{4}B_{1}+B_{2}B_{4}B_{2})\}=0
\end{eqnarray}

\begin{eqnarray} \label{c2}  4\{B_{3},B_{4}\}=\{(B_{1}B_{3}B_{1}+B_{2}B_{3}B_{2}),(B_{1}B_{4}B_{1}+B_{2}B_{4}B_{2})\}
\end{eqnarray}

\begin{eqnarray} \label{c3} && 4\{B_{3},B_{4}\}-2\{B_{3},(B_{1}B_{4}B_{1}+B_{2}B_{4}B_{2})\}-2\{B_{4},(B_{1}B_{3}B_{1}+B_{2}B_{3}B_{2})\}\nonumber \\
    && \hspace{0.3cm}+\{(B_{1}B_{3}B_{1}+B_{2}B_{3}B_{2}),(B_{1}B_{4}B_{1}+B_{2}B_{4}B_{2})\}=0 \nonumber\\
&& \Rightarrow 4\{B_{3},B_{4}\}=\{B_{3},(B_{1}B_{4}B_{1}+B_{2}B_{4}B_{2})\} +
\{B_{4},(B_{1}B_{3}B_{1}+B_{2}B_{3}B_{2})\} \ \ (\text{from Eq.}\ [\ref{c2})]
\end{eqnarray}

Putting Eqs. (\ref{c2}) and (\ref{c3}) in Eq. (\ref{c1}) and rearranging, again we get the following conditions
\begin{eqnarray} \label{c4}
   && \{B_{3},B_{4}\}=0\\
   \nonumber
   &&\{B_{3},(B_{1}B_{4}B_{1}+B_{2}B_{4}B_{2})\}+\{B_{4},(B_{1}B_{3}B_{1}+B_{2}B_{3}B_{2})\}=0\\
   \nonumber
    &&\{(B_{1}B_{3}B_{1}+B_{2}B_{3}B_{2}),(B_{1}B_{4}B_{1}+B_{2}B_{4}B_{2})\}=0.
\end{eqnarray}

 It is then straightforward to show that such anti-commuting  relation implies $B_{1}=B_{3}$ and $B_{2}=B_{4}$, and in turn $\{B_{1},B_{2}\}=0$. Hence, for optimal sequential quantum violations the set of observables for Bob$_{1}$ and Bob$_{2}$ is the same.

\section{Evaluation of quantum value of the CHSH expression $\langle \mathcal{B}^{2} \rangle^{g=4}_Q$ for Alice and Bob$_{2}$}
	\label{appg4}
		The  $g$-POVMs for set of $B_{y_{1}}$ are constructed as follows,
		\begin{eqnarray}
			\label{ag4}
			\nonumber
			E_{+|y_{1}}^{1}&=& \dfrac{1}{4} \left[ \mathbb{I} + \lambda_{1}\left( B_{y_{1}}+M_{y_{1}}\right)\right];\ 
			E_{+|y_{1}}^{2}= \dfrac{1}{4} \left[ \mathbb{I} +  \lambda_{1}\left( B_{y_{1}}-M_{y_{1}}\right)\right]\\
			E_{-|y_{1}}^{1}&=& \dfrac{1}{4} \left[ \mathbb{I} - \lambda_{1} \left( B_{y_{1}}+N_{y_{1}}\right)\right]; \  
			E_{-|y_{1}}^{2}= \dfrac{1}{4} \left[ \mathbb{I} -  \lambda_{1}\left( B_{y_{1}}-N_{y_{1}}\right)\right]
		\end{eqnarray} 
		where $M_{y_{1}}=B_{y_{1}}^{1}+ B_{y_{1}}^{2}$ and $N_{y_{1}}=B_{y_{1}}^{1}- B_{y_{1}}^{2}$.
		Here,  $\sum_{b_{k},j}E_{b_{k}|y_{1}}^{j}=\sum_{b_{k},j} (K_{b_{k}|y_{1}}^{j})^\dagger K_{b_{k}|y_{1}}^{j}=\mathbb{I}$ and $ \sqrt{E_{b_{k}|y_{1}}^{j} }=K_{b_{k}|y_{1}}^{j}$.
			Assuming $\lambda_{1}=1$, the $g$-POVMs in Eq. (\ref{ag4}) provides the projectors for $g=4$. To satisfy the  orthogonality criteria, Bobs observables have to satisfy the conditions given as 
		\begin{equation}
			[B_{y_{1}},B_{y_{1}}^{1}]=0, \ \ \ [B_{y_{1}}^{1},B_{y_{1}}^{2}]=0.
		\end{equation}
		
		\subsection{Construction of the Kraus operators}
	
		To construct the Kraus operators,  first using the property $\sum_{i=1}^{4}P_{i}=\mathbb{I}$, from Eq. (\ref{ag4})  we construct the elements of POVM where $P_{i}$ are the projectors corresponding to each of the eigenvectors from the set that constitutes the common eigen basis of $B_{y_{1}}$.  Then  we can write
\begin{eqnarray}
E_{+|{y_{1}}}^{1}&=&\dfrac{1}{4}\Big[ \mathbb{I}+\lambda_{1} \Big( B_{y_1}+ M_{y_{1}}\Big)\Big]=\dfrac{1}{4}\Big[ \mathbb{I}+\lambda_{1} \Big( 4P_{1}-\mathbb{I}\Big)\Big] \ ; \  \text{where} \  P_{1}=\frac{1}{4} \qty[\mathbb{I}+ \qty(B_{y_1}+B_{y_1}^{1}+B_{y_1}^{2})]\nonumber\\
&=&\dfrac{1}{4}\Big[\Big(1-\lambda_{1}\Big)\mathbb{I}+4\lambda_{1}P_{1}\Big]=\dfrac{1}{4}\Big[\Big(1-\lambda_{1}\Big)\Big(\mathbb{I}-P_{1}\Big)+\Big(1+3\lambda_{1}\Big) P_{1}\Big].
\end{eqnarray}

		Corresponding Kraus operator can be written as
		\begin{eqnarray}
			\nonumber
			K_{+|{y_{1}}}^{1}&=&\sqrt{\dfrac{1-\lambda_{1}}{4}}\mathbb{I}+ \Bigg(\sqrt{\dfrac{1+3\lambda_{1}}{4}}-\sqrt{\dfrac{1-\lambda_{1}}{4}} \Bigg)P_{1}\\
			\nonumber
			&=&\sqrt{\dfrac{1-\lambda_{1}}{4}} \mathbb{I}+\dfrac{1}{4}\Bigg(\sqrt{\dfrac{1+3\lambda_{1}}{4}}-\sqrt{\dfrac{1-\lambda_{1}}{4}} \Bigg)\Big(\mathbb{I}+ B_{y_1}+ M_{y_{1}}\Big)\\
			\nonumber
			&=&\dfrac{1}{4}\Bigg(\sqrt{\dfrac{1+3\lambda_{1}}{4}}+3\sqrt{\dfrac{1-\lambda_{1}}{4}} \Bigg)\mathbb{I}+\dfrac{1}{4}\Bigg(\sqrt{\dfrac{1+3\lambda_{1}}{4}}-\sqrt{\dfrac{1-\lambda_{1}}{4}} \Bigg)\Big(B_{y_1}+ M_{y_{1}}\Big)\\
			&=&\alpha_{4} \mathbb{I} + \beta_{4} X_{+|y_{1}}^{1} 
		\end{eqnarray}
where, $\alpha_4 = \dfrac{1}{4} \left[ \sqrt{\frac{1+3\lambda_{1} }{4}} + 3 \sqrt{\frac{1-\lambda_{1} }{4}}\right];\beta_4 = \dfrac{1}{4} \left[ \sqrt{\frac{1+3\lambda_{1} }{4}} - \sqrt{\frac{1-\lambda_{1} }{4}}\right]$ and $X_{+|y_{1}}^{1}=\Big(B_{y_1}+ M_{y_{1}}\Big)$. Similarly, other Kraus operators can also be constructed. Thus, the general Kraus operators can be represented in a simpler form as,	$K_{b_{k}|y_{1}}^{j}= \alpha_{4} \mathbb{I} + \beta_{4} X_{b_{k}|y_{1}}^{j} $ with $j\in \{1,2\}$.

	\subsection{Derivation of the quantum value of $\langle \mathcal{B}^{2} \rangle^{g=4}_Q$ for Alice and Bob$_{2}$}	
		The shared state between Alice and Bob$_{2}$ after the unsharp measurement of Bob$_{1}$ will be
\begin{eqnarray}
\rho_{AB_{2}}^{g=4}&=&\dfrac{1}{2}\sum_{j=1}^{2}\sum_{y_{1}=1}^{2} \sum_{b_{1}=\pm}\qty(\mathbb{I} \otimes K_{b_{1}|y_{1}}^{j})  \rho_{AB_{1}}^{g=4}  \left(\mathbb{I} \otimes K_{b_{1}|y_{1}}^{j}\right)\nonumber\\
&=&\dfrac{1}{2}\bigg[K_{+|1}^{1}\rho_{AB_{1}}^{g=4}K_{+|1}^{1} +K_{+|1}^{2}\rho_{AB_{1}}^{g=4}K_{+|1}^{2} + K_{-|1}^{1}\rho_{AB_{1}}^{g=4}K_{-|1}^{1}+K_{-|1}^{2}\rho_{AB_{1}}^{g=4}K_{-|1}^{2}+K_{+|2}^{1}\rho_{AB_{1}}^{g=4}K_{+|2}^{1} +K_{+|2}^{2}\rho_{AB_{1}}^{g=4}K_{+|2}^{2} + K_{-|2}^{1}\rho_{AB_{1}}^{g=4}K_{-|2}^{1}+K_{-|2}^{2}\rho_{AB_{1}}^{g=4}K_{-|2}^{2}\bigg]\nonumber\\
&=&  4 \alpha_{4}^2 (\mathbb{I}\otimes\mathbb{I})\rho_{AB_{1}}^{g=4}(\mathbb{I}\otimes\mathbb{I}) + 2\beta_{4}^2\Big[\left( (\mathbb{I}\otimes B_{1})  \rho_{AB_{1}}^{g=4} (\mathbb{I}\otimes B_{1}) + (\mathbb{I}\otimes B_{2}) \rho_{AB_{1}}^{g=4} (\mathbb{I}\otimes B_{2})\right) \nonumber\\
&+&\left((\mathbb{I}\otimes B_{1}^{1})  \rho_{AB_{1}}^{g=4} (\mathbb{I}\otimes B_{1}^{1}) + (\mathbb{I}\otimes B_{2}^{1}) \rho_{AB_{1}}^{g=4} (\mathbb{I}\otimes B_{2}^{1}) \right) + \left((\mathbb{I}\otimes B_{1}^{2})  \rho_{AB_{1}}^{g=4} (\mathbb{I}\otimes B_{1}^{2}) + (\mathbb{I}\otimes B_{2}^{2}) \rho_{AB_{1}}^{g=4} (\mathbb{I}\otimes B_{2}^{2})\right)\Big]\\
&=&  4 \alpha_{4}^{2} \ \rho_{AB_{1}}^{g=4} +2\beta_{4}^{2} \ \Bigg[\sum_{y_{1}=1}^{2} (\mathbb{I} \otimes B_{y_{1}}) \rho_{AB_{1}}^{g=4} (\mathbb{I} \otimes B_{y_{1}})+\sum_{{y_{1}}=1}^{2} \sum_{j=1}^{2}(\mathbb{I} \otimes B_{y_{1}}^{j}) \rho_{AB_{1}}^{g=4} (\mathbb{I} \otimes  B_{y_{1}}^{j}) \Bigg] 
\end{eqnarray} 
where $B_{y_{1}}$, $B_{y_{1}}^{1}$ and $B_{y_{1}}^{2}$ are mutually commuting observables.
		
The expectation value of the CHSH expression between Alice and Bob$_{2}$ for degeneracy breaking measurement is given by
\begin{equation}
\langle \mathcal{B}^{2} \rangle^{g=4}_Q=\Tr[\rho_{AB_{2}}^{g=4} \mathcal{B}]= 4(\alpha_{4}^2 + \beta_{4}^2)\langle \mathcal{B} \rangle_Q^{opt}
\end{equation}
where following the similar SOS approach as derived in Appendix \ref{appsos2}, it is straightforward to show that $B_{1}$ and $B_{2}$ have to be mutually anti-commuting and $\{ B_{1}, B_{1}^{1}, B_{1}^{2} \} $ ($\{ B_{2}, B_{2}^{1}, B_{2}^{2} \} $) have to be mutually commuting to obtain the maximum quantum value of $\langle \mathcal{B}^{2} \rangle^{g=4}_Q$.

\subsection{Generalized example for $g=4$ case}
Horodecki et. al. \cite{horodecki} have provided the construction of  the general representation of a two-qubit state (mixed or pure) in terms of Pauli matrices. For our case we are considering a particular observable, $B_{y_{1}}=\sigma_{z} \otimes \sigma_{z}$ which can be decomposed as $B_{y_{1}}=\ket{00}\bra{00}+\ket{00}\bra{11}+\ket{11}\bra{00}+\ket{11}\bra{11}$. The eigenstates for $B_{y_{1}}$ are the four Bell states given as $\{\ket{\psi_{+}},\ket{\psi_{-}},\ket{\phi_{+}},\ket{\phi_{-}}\}$. This choice of basis for $B_{y_{1}}$ are not unique and can be written in infinitely many ways. Let us choose a general basis for $B_{y_{1}}$ as $\ket{\gamma_{1}}=\xi\ket{00}+\sqrt{1-\xi^{2}}\ket{11}$, $\ket{\gamma_{2}}=\sqrt{1-\xi^{2}}\ket{00}-\xi\ket{11}$, $\ket{\gamma_{3}}=\xi\ket{01}+\sqrt{1-\xi^{2}}\ket{10}$ and $\ket{\gamma_{4}}=\sqrt{1-\xi^{2}}\ket{01}-\xi\ket{10}$. In this case, for $g=4$ the generalized $g$-POVMs for $B_{y_{1}}$ where $y_{1}=1,2$ is evaluated as
		\begin{eqnarray}
			\label{povm3q}
			\nonumber
			E_{+|y_{1}}^{1}&=& \dfrac{1}{8} \left[ \mathbb{I} + \lambda \left( B_{y_{1}}+ (2\xi^{2}-1)B_{y_{1}}^{1}+(2\xi^{2}-1)B_{y_{1}}^{2} + 2\xi\sqrt{1-\xi^{2}}B_{y_{1}}^{3}- 2\xi\sqrt{1-\xi^{2}}B_{y_{1}}^{4} \right)\right]\\
			\nonumber
			E_{+|y_{1}}^{2}&=& \dfrac{1}{8} \left[ \mathbb{I} + \lambda \left( B_{y_{1}}+ (1-2\xi^{2})B_{y_{1}}^{1}+(1-2\xi^{2})B_{y_{1}}^{2} - 2\xi\sqrt{1-\xi^{2}}B_{y_{1}}^{3}+ 2\xi\sqrt{1-\xi^{2}}B_{y_{1}}^{4}\right)\right]\\
			\nonumber
			E_{-|y_{1}}^{1}&=& \dfrac{1}{8} \left[ \mathbb{I} - \lambda \left(B_{y_{1}}- (1-2\xi^{2})B_{y_{1}}^{1}-(2\xi^{2}-1)B_{y_{1}}^{2} - 2\xi\sqrt{1-\xi^{2}}B_{y_{1}}^{3}- 2\xi\sqrt{1-\xi^{2}}B_{y_{1}}^{4} \right)\right]\\
			E_{-|y_{1}}^{2}&=& \dfrac{1}{8} \left[ \mathbb{I} - \lambda \left( B_{y_{1}}- (2\xi^{2}-1)B_{y_{1}}^{1}-(1-2\xi^{2})B_{y_{1}}^{2} + 2\xi\sqrt{1-\xi^{2}}B_{y_{1}}^{3}+ 2\xi\sqrt{1-\xi^{2}}B_{y_{1}}^{4} \right)\right]
		\end{eqnarray}
		where $0<\xi<1$. 
		
Given any value of $\xi$, the quantum value of the CHSH expression between Alice and Bob$_{2}$, $\langle \mathcal{B}^{2} \rangle^{g=4}_Q$ remains intact i.e., $\langle \mathcal{B}^{2} \rangle^{g=4}_Q$ values for $g$-POVM measurements are independent of $\xi$.
	\section{Detailed calculation for sequential quantum value of the CHSH expression $\langle \mathcal{B}^{2} \rangle^{g=3}_Q$ for Alice and Bob$_{2}$}
	\label{appg3}
		In the case of $g=3$ POVM measurement, we can construct a rank-2 POVM by adding two rank-1 POVMs as $E_{+|y_{1}}^{12}=E_{+|y_{1}}^{1}+E_{+|y_{1}}^{2}=\dfrac{1}{2}(\mathbb{I}+\lambda_{1} B_{y_{1}})$. The Kraus operator corresponding to the POVMs are given as $K_{+|y_{1}}^{12}= (\alpha_{2} \mathbb{I} +\beta_{2} B_{y_{1}}).$
		The other two rank-1 POVMs are defined as $K_{-|y_{1}}^{1}=(\alpha_{4} \mathbb{I}+\beta_{4} X_{-|y_{1}}^{1})$ and $K_{-|y_{1}}^{2}=(\alpha_{4} \mathbb{I}+\beta_{4} X_{-|y_{1}}^{2})$. 
		
		Therefore, shared state between Alice and Bob$_{2}$ after the unsharp measurement of Bob$_{1}$ using above 3 POVMs will be
\begin{eqnarray}
\rho^{g=3}_{AB_{2}}&=&\frac{1}{2} \sum_{y_{1}=1}^{2}\qty[\qty( \mathbb{I} \otimes K_{+|y_{1}}^{12}) \rho_{AB_{1}}^{g=3}\qty( \mathbb{I} \otimes K_{+|y_{1}}^{12}) + \qty( \mathbb{I} \otimes K_{-|y_{1}}^{1}) \rho_{AB_{1}}^{g=3} \qty(\mathbb{I} \otimes K_{-|y_{1}}^{1})+ \qty(\mathbb{I} \otimes K_{-|y_{1}}^{2})\rho_{AB_{1}}^{g=3}\qty( \mathbb{I} \otimes K_{-|y_{1}}^{2})]\nonumber\\
&=&\frac{1}{2} \Bigg[ \Big( \alpha_{2} \mathbb{I} \otimes \mathbb{I} + \beta_{2} \mathbb{I} \otimes B_{1}\Big) \rho_{AB_{1}}^{g=3} \Big( \alpha_{2} \mathbb{I} \otimes \mathbb{I} + \beta_{2} \mathbb{I} \otimes B_{1}\Big) + \Big(\alpha_{4} \mathbb{I} \otimes \mathbb{I}+ \beta_{4} \mathbb{I} \otimes X_{-|1}^{1}\Big)\rho_{AB_{1}}^{g=3} \Big(\alpha_{4} \mathbb{I} \otimes \mathbb{I}+ \beta_{4} \mathbb{I} \otimes X_{-|1}^{1}\Big)\nonumber\\
&+&\Big(\alpha_{4} \mathbb{I} \otimes \mathbb{I}+ \beta_{4} \mathbb{I} \otimes X_{-|1}^{2}\Big)\rho_{AB_{1}}^{g=3} \Big(\alpha_{4} \mathbb{I} \otimes \mathbb{I}+ \beta_{4} \mathbb{I} \otimes X_{-|1}^{2}\Big)+\Big( \alpha_{2} \mathbb{I} \otimes \mathbb{I} + \beta_{2} \mathbb{I} \otimes B_{2}\Big) \rho_{AB_{1}}^{g=3} \Big( \alpha_{2} \mathbb{I} \otimes \mathbb{I} + \beta_{2} \mathbb{I} \otimes B_{2}\Big)\\
&+& \Big(\alpha_{4} \mathbb{I} \otimes \mathbb{I}+ \beta_{4} \mathbb{I} \otimes X_{-|2}^{1}\Big)\rho_{AB_{1}}^{g=3} \Big(\alpha_{4}\mathbb{I} \otimes  \mathbb{I}+ \beta_{4} \mathbb{I} \otimes X_{-|2}^{1}\Big)+\Big(\alpha_{4} \mathbb{I} \otimes \mathbb{I}+ \beta_{4} \mathbb{I} \otimes X_{-|2}^{2}\Big)\rho_{AB_{1}}^{g=3} \Big(\alpha_{4} \mathbb{I} \otimes \mathbb{I}+ \beta_{4} \mathbb{I} \otimes X_{-|2}^{2}\Big)\Bigg]
\end{eqnarray}
The quantum value of the CHSH expression between Alice and Bob$_{2}$ for $g=3$-POVM measurement is calculated as
\begin{equation}
\langle \mathcal{B}^{2} \rangle^{g=3}_Q=\Tr[\rho_{AB_{2}}^{g=3}\mathcal{B}]= \qty[\alpha_{2}^2 + 2 \Big( \alpha_{4}^2 + \beta_{4}^2\Big)] \langle \mathcal{B} \rangle_Q^{opt}
\end{equation} 
	
\section{Detailed calculation for sequential quantum value of the CHSH expression for Higher degree of degeneracy-breaking measurement}\label{appg}
Bob$_{1}$ can further extend the possibility of degeneracy-breaking measurement in dimension $d=2^{ m}$ where $m$ is arbitrary considering higher number of $g$-POVMs.  Let us take, $g=8$ case as an example where the $g$-POVMs are explicitly defined as
\begin{eqnarray}
E_{+|y_{1}}^{1}&=& \dfrac{1}{8} \left[ \mathbb{I} + \lambda \left( B_{y_{1}}+M_{y_{1}}^{1}+ M_{y_{1}}^{2}+M_{y_{1}}^{3}\right)\right];\ \
E_{+|y_{1}}^{2}= \dfrac{1}{8} \left[ \mathbb{I} + \lambda \left( B_{y_{1}}+M_{y_{1}}^{1}- M_{y_{1}}^{2}-M_{y_{1}}^{3}\right)\right];\nonumber\\
E_{+|y_{1}}^{3}&=& \dfrac{1}{8} \left[ \mathbb{I} + \lambda \left( B_{y_{1}}-M_{y_{1}}^{1}- M_{y_{1}}^{2}+M_{y_{1}}^{3}\right)\right];\ \
E_{+|y}^{4}= \dfrac{1}{8} \left[ \mathbb{I} + \lambda \left( B_{y_{1}}-M_{y_{1}}^{1}+ M_{y_{1}}^{2}-M_{y_{1}}^{3}\right)\right];\nonumber\\
E_{-|y_{1}}^{1}&=& \dfrac{1}{8} \left[ \mathbb{I} - \lambda \left( B_{y_{1}}-N_{y_{1}}^{1}- N_{y_{1}}^{2}-N_{y_{1}}^{3}\right)\right];\ \
E_{-|y_{1}}^{2}=\dfrac{1}{8} \left[ \mathbb{I} - \lambda \left( B_{y_{1}}-N_{y_{1}}^{1}+ N_{y_{1}}^{2}+N_{y_{1}}^{3}\right)\right];\nonumber\\
E_{-|y_{1}}^{3}&=& \dfrac{1}{8} \left[ \mathbb{I} -\lambda \left( B_{y_{1}}+N_{y_{1}}^{1}- N_{y_{1}}^{3}+N_{y_{1}}^{3}\right)\right];\ \
E_{-|y_{1}}^{4}=\dfrac{1}{8} \left[ \mathbb{I} - \lambda \left( B_{y_{1}}+N_{y_{1}}^{1}+ N_{y_{1}}^{3}-N_{y_{1}}^{3}\right)\right];
\end{eqnarray}
with $\sum_{j=1}^{4} E_{b_{k}|y_{1}}^{j}=\mathbb{I}$ and $M_{y_{1}}^{i}$($N_{y_{1}}^{i}$) having $i \in \{1,2,3\}$ are explicitly written as
$M_{y_{1}}^{1}=  \left( B_{y_{1}}^{1}+ B_{y_{1}}^{2}\right)$, $M_{y_{1}}^{2}= \left( B_{y_{1}}^{3}+B_{y_{1}}^{4}\right)$, $M_{y_{1}}^{3}= \left( B_{y_{1}}^{5}+ B_{y_{1}}^{6}\right)$, $N_{y_{1}}^{1}= \left( B_{y_{1}}^{1}-B_{y_{1}}^{2}\right)$, $N_{y_{1}}^{2}=\left( B_{y_{1}}^{3}-B_{y_{1}}^{4}\right)$, and $N_{y_{1}}^{3}=\left( B_{y_{1}}^{5}-B_{y_{1}}^{6}\right)$. $M_{y_{1}}^{i}$ and $N_{y_{1}}^{i}$ are not unique but will not effect the final results.
		
The shared reduced state between Alice and Bob$_{2}$ after the non-selective measurement of Bob$_{1}$ can be written as
\begin{eqnarray}\label{dbg8}
\rho_{AB_{2}}^{g=8}&=&\dfrac{1}{2}\sum_{y_{1}=1}^{2} \sum_{j=1}^{4} \sum_{b_{1}\in \pm} \left(\mathbb{I} \otimes K_{b_{1}|{y_{1}}}^{j}\right)  \rho_{AB_{1}}^{g=8}  \left(\mathbb{I} \otimes K_{b_{1}|{y_{1}}}^{j}\right)\nonumber\\
&=&8\alpha_8^2 \rho_{AB_{1}}^{g=8} +4 \beta_8^2\sum_{y_{1}=1}^{2} \Bigg[(\mathbb{I} \otimes B_{y_{1}}) \rho_{AB_{1}}^{g=8} (\mathbb{I} \otimes B_{y_{1}})+ \sum_{i=1}^{3}\Big((\mathbb{I} \otimes M_{y_{1}}^{i}) \rho_{AB_{1}}^{g=8} (\mathbb{I} \otimes  M_{y_{1}}^{i})+(\mathbb{I} \otimes N_{y_{1}}^{i}) \rho_{AB_{1}}^{g=8} (\mathbb{I} \otimes  N_{y_{1}}^{i})\Big)\Bigg] \end{eqnarray}
where $K^{j}_{b_{1}|y_{1}}=\sqrt{E_{b_{1}|y_{1}}^{j}}$ are Kraus operators which can be written into a simpler form as	$K_{b_{k}|y_{1}}^{j}= \alpha_{8} \mathbb{I} + \beta_{8} X_{b_{k}|y_{1}}^{j} $ having
\begin{equation}
\alpha_8 =\frac{1}{8} \left[ \sqrt{\frac{1+7\lambda}{8}} + 7 \sqrt{\frac{1-\lambda}{8}}\right];\ \ \
\beta_8 =\frac{1}{8} \left[ \sqrt{\frac{1+7\lambda}{8}} - \sqrt{\frac{1-\lambda}{8}}\right].
\end{equation} 
Replacing $K_{b_{k}|y_{1}}^{j}=(\alpha_{8} \mathbb{I}\otimes \mathbb{I}+ \beta_{8} \mathbb{I} \otimes X_{b_{k}|y_{1}}^{j})$ in the Eq. (\ref{dbg8}), we get
\begin{eqnarray}
\rho_{AB_{2}}^{g=8}&=&  8 \alpha_{8}^2 \qty(\mathbb{I}\otimes\mathbb{I})\rho_{AB_{1}}^{g=8}\qty(\mathbb{I}\otimes\mathbb{I}) + 4\beta_{8}^2\Big[ \qty(\mathbb{I}\otimes B_{1})  \rho_{AB_{1}}^{g=8} \qty(\mathbb{I}\otimes B_{1}) + \qty(\mathbb{I}\otimes B_{2}) \rho_{AB_{1}}^{g=8} \qty(\mathbb{I}\otimes B_{2})+(\mathbb{I}\otimes M_{1}^{1})  \rho_{AB_{1}}^{g=8} (\mathbb{I}\otimes M_{1}^{1}) + (\mathbb{I}\otimes M_{2}^{1}) \rho_{AB_{1}}^{g=8} (\mathbb{I}\otimes M_{2}^{1}) \nonumber\\
&+&(\mathbb{I}\otimes M_{1}^{2})  \rho_{AB_{1}}^{g=8} (\mathbb{I}\otimes M_{1}^{2}) + (\mathbb{I}\otimes M_{2}^{2}) \rho_{AB_{1}}^{g=8} (\mathbb{I}\otimes M_{2}^{2})+(\mathbb{I}\otimes M_{1}^{3})  \rho_{AB_{1}}^{g=8} (\mathbb{I}\otimes M_{1}^{3}) + (\mathbb{I}\otimes M_{2}^{3}) \rho_{AB_{1}}^{g=8} (\mathbb{I}\otimes M_{2}^{3})+(\mathbb{I}\otimes N_{1}^{1})  \rho_{AB_{1}}^{g=8} (\mathbb{I}\otimes N_{1}^{1})\nonumber\\
&+& (\mathbb{I}\otimes N_{2}^{1}) \rho_{AB_{1}}^{g=8} (\mathbb{I}\otimes N_{2}^{1})+(\mathbb{I}\otimes N_{1}^{2})  \rho_{AB_{1}}^{g=8} (\mathbb{I}\otimes N_{1}^{2})+ (\mathbb{I}\otimes N_{2}^{2}) \rho_{AB_{1}}^{g=8} (\mathbb{I}\otimes N_{2}^{2})+(\mathbb{I}\otimes N_{1}^{3})  \rho_{AB_{1}}^{g=8} (\mathbb{I}\otimes N_{1}^{3})+ (\mathbb{I}\otimes N_{2}^{3}) \rho_{AB_{1}}^{g=8} (\mathbb{I}\otimes N_{2}^{3})\Big]\nonumber\\
&=&8\alpha_8^2 \rho_{AB_{1}}^{g=8} +4 \beta_8^2\sum_{y_{1}=1}^{2} \qty[\qty(\mathbb{I} \otimes B_{y_{1}}) \rho_{AB_{1}}^{g=8} \qty(\mathbb{I} \otimes B_{y_{1}})+\sum_{i=1}^{3}\qty(\mathbb{I} \otimes M_{y_{1}}^{i}) \rho_{AB_{1}}^{g=8} \qty(\mathbb{I} \otimes  M_{y_{1}}^{i})+\qty(\mathbb{I} \otimes N_{y_{1}}^{i}) \rho_{AB_{1}}^{g=8} \qty(\mathbb{I} \otimes  N_{y_{1}}^{i})].
\end{eqnarray} 
The quantum value of the CHSH expression Alice and Bob$_{2}$ is derived as 
\begin{equation}
\langle \mathcal{B}^{2} \rangle^{g=8}_Q= \Tr[\rho_{AB_{2}}^{g=8} \mathcal{B}]=8 \ \qty(\alpha_8^2 + \beta_8^2 ) \langle \mathcal{B} \rangle_Q^{opt}.
\end{equation}

In terms of the unsharpness parameter we have
\begin{equation}
\langle \mathcal{B}^{2} \rangle^{g=8}_Q=\dfrac{1}{8}\qty[\qty(\dfrac{(13-9\lambda_{1})}{2}+\dfrac{3}{2}\sqrt{(1+7\lambda_{1})(1-\lambda_{1})})]\langle \mathcal{B} \rangle_Q^{opt}.
\end{equation}
Provided the critical value of unsharpness parameter $\lambda_{1}^{*}=1/\sqrt{2}$, the CHSH expression will not be violated for Alice and Bob$_{2}$ and non-locality cannot be shared up to two Bobs.
Let us now consider that the measurement is implemented by using $7$-POVMs. In that case, the relevant POVMs for $g=7$ can be construct as \{$E_{+|y_{1}}^{12}$,  $E_{+|y_{1}}^{3}$, $E_{+|y_{1}}^{4}$, $E_{-|y_{1}}^{1}$, $E_{-|y_{1}}^{2}$, $E_{-|y_{1}}^{3}$, $E_{-|y_{1}}^{4}$\} where $E_{+|y_{1}}^{12}\equiv E_{+|y_{1}}=E_{+|y_{1}}^{1}+E_{+|y_{1}}^{2}$. There will be many permutations through which one can implement $7$-POVM measurements. However, the results will remain unchanged. The quantum value of the CHSH expression between Alice and Bob$_{2}$ is calculated as
\begin{eqnarray}
\langle \mathcal{B}^{2} \rangle^{g=7}_Q& =&\Big[\alpha_{4} ^2+ 6(\alpha_{8}^2 + \beta_{8}^2 )\Big] \langle \mathcal{B} \rangle_Q^{opt}\nonumber\\
&=&\dfrac{1}{64}\Big[\Big(49-33\lambda_{1}\Big)+ 9\sqrt{\Big(1+7\lambda_{1} \Big)\Big( 1-\lambda_{1}\Big)}+ 6\sqrt{\Big(1+3\lambda_{1} \Big)\Big( 1-\lambda_{1}\Big)}\Big] \langle \mathcal{B} \rangle_Q^{opt}.
\end{eqnarray} 

Similarly, the quantum value of the CHSH expression between Alice and Bob$_{2}$ for $g=6$ POVMs \{$E_{+|y_{1}}^{12}$, $E_{+|y_{1}}^{34}$, $E_{-|y_{1}}^{1}$, $E_{-|y_{1}}^{2}$, $E_{-|y_{1}}^{3}$, $E_{-|y_{1}}^{4}$\} where $E_{+|y_{1}}^{12}=(E_{+|y_{1}}^{1}+E_{+|y_{1}}^{2})$, $E_{+|y_{1}}^{34}=(E_{+|y_{1}}^{3}+E_{+|y_{1}}^{4})$ can be computed as
\begin{eqnarray}
\langle \mathcal{B}^{2} \rangle^{g=6}_Q& =&\qty[2\alpha_{4} ^2+ 4\qty(\alpha_{8}^2 + \beta_{8}^2 )] \langle \mathcal{B} \rangle_Q^{opt}\nonumber\\
&=&\dfrac{1}{16}\qty[\qty(\dfrac{23-15\lambda_{1}}{2})+ \frac{3}{2}\sqrt{\qty(1+7\lambda_{1})\qty( 1-\lambda_{1})}+ 3\sqrt{\qty(1+3\lambda_{1})\qty( 1-\lambda_{1})}]\langle \mathcal{B} \rangle_Q^{opt}.
\end{eqnarray} 
where the upper bound of $\lambda_{1}$ is calculated as $(\lambda_{1})_{\max} \approx 0.686$ when Bob$_{2}$ performs sharp measurement.
		
Similarly, for $g=5$ case we get \{$E_{+|y_{1}}^{1234}$, $E_{-|y_{1}}^{1}$, $E_{-|y_{1}}^{2}$, $E_{-|y_{1}}^{3}$, $E_{-|y_{1}}^{4}$\} or vice-versa where $E_{+|y_{1}}^{1234}\equiv E_{+|y_{1}}=(E_{+|y_{1}}^{1}+E_{+|y_{1}}^{2}+E_{+|y_{1}}^{3}+E_{+|y_{1}}^{4})$ is a rank- 4 POVM which is same as in case of $g=2$ and the rest four are rank- 1 POVMs. The quantum value of the CHSH expression between Alice and Bob$_{2}$ can be calculated as
\begin{eqnarray}
\langle \mathcal{B}^{2} \rangle^{g=5}_Q&=&\qty[\alpha_{2} ^2+ 4\qty(\alpha_{8}^2 + \beta_{8}^2)]\langle \mathcal{B} \rangle_Q^{opt}\nonumber\\
&=&\dfrac{1}{16}\qty[ \qty(\dfrac{13-9\lambda_{1}}{2})+ 4\qty( 1+ \sqrt{1-\lambda_{1}^2})+ \sqrt{\qty(1+7\lambda_{1} )\qty( 1-\lambda_{1})}] \langle \mathcal{B} \rangle_Q^{opt}.
\end{eqnarray} 
where the upper bound of $\lambda_{1}$ is calculated as $(\lambda_{1})_{\max} \approx 0.777$ when Bob$_{2}$ performs sharp measurement.	

It can be easily shown that given a value of $\lambda_{1}$, we have  $\langle \mathcal{B}^{2} \rangle^{g=5}_{Q}>\langle \mathcal{B}^{2} \rangle^{g=6}_{Q}>\langle \mathcal{B}^{2} \rangle^{g=7}_{Q}>\langle \mathcal{B}^{2} \rangle^{g=8}_{Q}$. This simply means that more the number of POVMs used by Bob$_{1}$ for his measurement, more the information is extracted from the system which eventually leads less quantum violation of the Bell inequality for Alice and Bob$_{2}$, as expected.
	
This scheme can further be generalized for any arbitrary $g$-POVM when $g_{\max}=d$ where $d$ is also arbitrary. The quantum value of the CHSH expression is derived as
\begin{eqnarray}
\langle \mathcal{B}^{2} \rangle^{g=g_{max}}_Q& =& \Tr[\rho_{AB_{2}}^{g}\mathcal{B}]=(g-d)\alpha_{d/2}^2+(2g-d) (\alpha_{d}^2 + \beta_{d}^2 ) \langle \mathcal{B} \rangle_Q \ \text{for} \ g>d/2,\\
& =& (g-d)\alpha_{d/4}^2+(2g-d/2)  (\alpha_{d/2}^2 + \beta_{d/2}^2 ) \langle \mathcal{B} \rangle_Q\ \text{for} \  g\leq d/2,
\end{eqnarray}
where,
\begin{eqnarray}\label{alpha}
\alpha_{d}=\frac{1}{d} \qty[ \sqrt{\dfrac{1+(d-1)\lambda_{1}}{d}} +(d-1) \sqrt{\dfrac{1-\lambda_{1}}{d}}] \ \ ; \ \ \beta_{d} =\dfrac{1}{d} \left[ \sqrt{\dfrac{1+(d-1)\lambda_{1}}{d}} - \sqrt{\dfrac{1-\lambda_{1}}{d}}\right] .
\end{eqnarray}

\end{widetext}

\end{document}